\documentclass[preprint,aps,prb,showpacs,groupedaddress]{revtex4-1}
\usepackage{times,xspace}
\usepackage{amsbsy,amssymb,amsmath,bm}
\usepackage{graphicx,color,epsfig,rotate}

\begin{document}
\title{
Strain-induced metal-insulator phase coexistence and stability in
perovskite manganites }

\author{T. F. Seman,$^1$ K. H. Ahn,$^1$ T. Lookman,$^2$  and A. R. Bishop$^2$}
\affiliation{$^1$Department of Physics, New Jersey Institute of Technology, Newark, New Jersey 07102, USA \\
$^2$Theoretical Division, Los Alamos National Laboratory, Los Alamos, New Mexico 87545, USA}


\begin{abstract}
We present a detailed study of a model for strain-induced
metal-insulator phase coexistence in perovskite manganites. Both
nanoscale and mesoscale inhomogeneities are self-consistently described using atomic
scale modes and their associated constraint equations.  We also  examine the stability of domain
configurations against uniform and nonuniform modifications of domain
walls. Our results show that the long range
interactions between  strain fields and the complex energy landscape with
multiple metastable states play  essential roles in stabilizing
metal-insulator phase coexistence, as observed in perovskite manganites.
We elaborate on the modes,
constraint equations, energies, and energy gradients that form the basis of our
simulation results.
\end{abstract}

\pacs{75.47.Gk, 64.70.K-, 75.60.Ch, 75.47.Lx}

\maketitle


\newpage

\section{Introduction}

Over the last several years, much attention has  focused on the
multiscale inhomogeneities observed in perovskite
manganites.~\cite{Salamon01} Unlike inhomogeneities  for some
other complex electronic systems, such as stripes in high-T$_c$ cuprates, the
coexistence of metallic and insulating phases within the same
crystals of manganites has been directly observed through various
high resolution local probes, such as dark field images
and scanning microscopy.~\cite{Uehara99,Fath99,Renner02}
Nanoscale inhomogeneities have also been implied  based on x-ray diffraction
results.~\cite{Kiryukhin04} Theories based on chemical randomness and electronic phase
separation have been proposed as a mechanism for such
inhomogeneities.~\cite{Moreo99,Burgy04} However, theories based on
chemical randomness alone assume an exact degeneracy between metallic and
insulating phases,~\cite{Burgy04} which ultimately leads to a
homogeneous phase.~\cite{Ahn04arxiv} Also, the effect of the Coulomb
interaction has not been incorporated adequately into the model describing
electronic phase separation.~\cite{Moreo99}

We  have previously proposed an {\it intrinsic} mechanism for phase
coexistence,~\cite{Ahn04} in which the interaction between strain fields plays an
important role, as speculated in earlier
literature.~\cite{Mathur01,Millis03,Bishop03} Specifically, our
model~\cite{Ahn04} includes intrinsic complexity of the energy landscape
and long range anisotropic interaction between strain fields, and
shows how such physics can give rise to multiscale inhomogeneities
observed in manganites. Our theoretical idea is supported by
experimental results.~\cite{Mathur04} For example, the observed large scale
inhomogeneity of the order of 10 $\mu$m without any observable chemical
inhomogeneity at a length scale of 0.5 $\mu$m or larger, suggests
 an intrinsic mechanism  for the phase coexistence.~\cite{Sarma04}
The lamellar morphology of coexisting phases observed in
manganites and the change of domain configurations upon
thermal cycling between 10 K and 300 K further support this point of view.~\cite{Tao05}
It is also found that the anisotropic epitaxial strain in thin films
gives rise to anisotropic percolation, which suggests that the origin of phase
coexistence is much more strongly affected by long range strain rather than by
local chemical inhomogeneity due to doping.~\cite{Ward09}

In contrast to the point of view in Refs.~\onlinecite{Moreo99} and~\onlinecite{Burgy04}
that considers {\it extrinsic} mechanism such as chemical inhomogeneity and disorder
as being key to understanding coexistence of metallic and insulating phases,
our work proposes that the {\it intrinsic} mechanism, that is, the competition
between the short and long range interactions, creates a delicate energy
landscape that leads to  domain formation. Further, the domain walls can be
pinned by the atomistic Peierls-Nabarro force, rather than by disorder
which is frequently attributed as being the cause for the
phenomenon.~\cite{Moreo99,Burgy04}
We proposed the basic ideas in
Ref.~\onlinecite{Ahn04}, but now develop them further and describe more extensive
implementation of the approach here.
Specifically, we present details of our model, methods, and
results, as well as further simulations on the stability of phase
coexistence. In Sect. II, the details of the Hamiltonian used for
the simulations in Ref.~\onlinecite{Ahn04}, and results obtained with
various initial conditions and parameters, are presented. We also
contrast these results with simulations for a system that include short
range interactions only. In Sect. III, we discuss the mechanism
underlying the stability of micron-scale phase coexistence through
further simulations and analysis. In Sect. IV, we summarize our main results.
Expressions for energies and energy gradients
used for simulations of the inhomogeneous states are provided in Appendices A and B,
respectively.

\section{Model for strain-induced metal-insulator phase coexistence}
\subsection{Properties of manganites and requirements  for phase coexistence in manganites}
Perovskite manganites typically have the chemical formula
$RE_{1-x}AK_x$MnO$_3$, where $RE$ represents rare earth elements,
such as La, Nd, and Pr, and $AK$ represents alkali metal elements
such as Ca and Sr.~\cite{Millis98,Salamon01,Mathur03} The important
electrons for both electronic properties and structures are the
$e_g$ electrons on Mn ions: the degeneracy of the $e_g$ orbitals leads to
a strong Jahn-Teller electron-lattice coupling. Shortly after the
discovery of colossal magnetoresistance in these
materials,~\cite{Jin94} the importance of strong electron-lattice
coupling was pointed out to explain the large resistivity above the
ferromagnetic transition temperature in terms of dynamic Jahn-Teller
polarons.~\cite{Millis98} The same strong electron-lattice coupling
is also responsible for a large structural difference between the
low temperature metallic and insulating phases of these materials.
In the insulating phase, the $e_g$ electrons are localized and the
charge density forms an ordered pattern. The orbital states of the
$e_g$ electrons, which are linear combinations of two $e_g$ orbital
states, $x^2-y^2$ and $3z^2-r^2$, are also ordered. Due to the
static Jahn-Teller effect, such a charge and orbital ordered state
accompanies uniform and short wavelength lattice distortions. For
example, the long Mn-O bond (or the elongated $e_g$ electron
orbital) in La$_{0.5}$Ca$_{0.5}$MnO$_3$ alternates its direction in
a plane, which gives rise to short wavelength lattice
distortions.~\cite{Chen96} Along the direction perpendicular to this
plane, the short Mn-O bond repeats itself, and therefore the unit cell
is compressed along this orientation, giving  uniform or
long-wavelength lattice distortions. In contrast, the lattice in the
metallic ground state of manganites has a structure close to an
ideal cubic perovskite structure without Jahn-Teller lattice
distortions because the electrons are delocalized. The ground state
can be changed between metallic and insulating phases in various ways, such as by the
size of $RE$ and $AK$, applied magnetic
fields, or applied pressures.~\cite{Hwang95,Tokura96,Hwang95b}

We propose that the first key to the understanding of the phase
coexistence in manganites is the metastability, which has been
observed in many experiments. For example, the distorted insulating
phase of manganites can be transformed into the undistorted metallic phase by
either x-rays~\cite{Kiryukhin97} or magnetic fields,~\cite{Tokura96}
and the metallic phase does not revert to the insulating phase
even after the external perturbation is removed. In particular, in
 x-ray experiments,~\cite{Kiryukhin97} the reduction of the superlattice peak
intensity and the simultaneous increase of conductivity, while the
sample is exposed to the x-rays, demonstrate the transformation of the
insulating phase into the metallic phase and the presence of inhomogeneity. However,
an energy landscape with local and global energy minima is not sufficient to explain the
observed sub-micron scale inhomogeneity, because such inhomogeneity
is unstable against thermal fluctuations. As pointed out in
Ref.~\onlinecite{Mathur03}, an unusual aspect of inhomogeneity in
manganites is its stability over a 100 K range in temperature, which is an indication of an extra mechanism at play that affects
phase coexistence in manganites.

Based on the strong electron-lattice coupling mentioned above and experiments showing the  important role
of strain in metal-insulator transition in manganites,~\cite{Podzorov01}
we propose that the
extra mechanism should be related to the long range anisotropic interaction
between {\it strain} fields. It is thus essential to consider the energy
landscape in terms of the lattice distortion variables, which will ultimately have a bearing on
other degrees of freedom such as magnetic moment or electron
density. The origin of the long range
anisotropic interaction within this framework is the bonding constraint, often
referred to as strain compatibility. The compatibility condition enforces
single-valued strain fields without broken  bonds.~\cite{Shenoy99,Lookman03,Ahn03}
The anisotropy reflects the discrete rotational symmetry
associated with the lattice structures.
Such long-range anisotropic interactions are responsible, for example, for
well-defined  structural twin boundaries~\cite{Ren02} over distances of
100~$\mu m$.

The cubic phase of perovskite manganites contains five atoms per unit cell.
The insulating charge and orbital ordered phase consists of a zig-zag
pattern of the long Mn-O bond orientation, which further increases the
number of atoms per unit cell. Inclusion of such details
is necessary for a complete description of properties of these materials.
For the current study,
however, we wish to focus on the following three key features of manganites essential for
multiscale inhomogeneity, and capture them in a simple model.
{\it First}, the metallic phase has
almost no lattice distortions in comparison with the charge and
orbital ordered insulating phase. {\it Second}, the insulating ground
state has a uniform or long wavelength ($\vec{k} \sim 0$) lattice
distortion. This property is essential because it is the long
wavelength distortion, not the short wavelength one, that
gives rise to the long range anisotropic interaction between strain
fields. {\it Third}, the insulating phase has a short wavelength lattice
distortion, in addition to the uniform distortion. As we will show below,
the symmetry-allowed coupling between uniform and
short-wavelength distortions gives rise naturally to an energy
landscape with multiple minima.

\subsection{Model system, variables and constraint equations}
Before we introduce our model, we examine whether a simpler 2-dimensional (2D) model can be used instead of
a 3-dimensional (3D) model, in particular to capture the effect of
the long-range strain-strain interactions. In D-dimensional space (D
= 2 or 3), the anisotropic strain-strain interaction decays as
$1/r^D$, where $r$ represents the distance between two
points.~\cite{Shenoy99,Rasmussen01, Lookman03}
The spatial integration of $1/r^D$ would give rise to a logarithmic divergence
in both 2 and 3 dimensional space,~\cite{Anisotropy}
which indicates that the effect of the interaction
would be similar for both cases. Indeed, recent
simulations of strains in 2 and 3 dimensional space show very
similar results.~\cite{Rasmussen01, Lookman03}
Thus, we limit ourselves in this work to  a 2D model for simplicity.

\begin{figure}[ht]
\includegraphics[scale=1.0, clip=true]{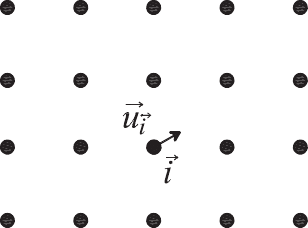}
\caption{\label{fig:square}
Two-dimensional square lattice with a monatomic basis.
$\vec{u}_{\vec{i}}$ represents the displacement of the atom at the
site with the index $\vec{i}=(i_x,i_y)$, where integer $i_x$ and $i_y$ range from 1 to $N$.
The site at the bottom left corner is chosen for $\vec{i}=(1,1)$.
In this work, the lattice constant $a$ is irrelevant for
the expressions of modes and energies, and can be chosen arbitrarily, so
long as the sequence of the atoms is not changed by the
displacements. For figures in this work,
$a$ is chosen as 10, but the result itself is independent of the choice of $a$.
}
\end{figure}

One of the simplest lattices in 2D space is the square lattice with a
monatomic basis shown in Fig.~\ref{fig:square}. By considering one
isotropic electron orbital per site and  nearest neighbor electron hopping,
the lattice supports a metallic electron density of states (DOS)
without a gap. Therefore, such an undistorted square lattice shares the
first property for manganites mentioned above. To include the second
property, we deform the square unit cell of the
lattice to a rectangular unit cell, either along the horizontal or
vertical directions. To include the third property,
 we incorporate the $(\pi,\pi)$ type displacements of atoms
along the horizontal (vertical) direction for a rectangular lattice
elongated along the vertical (horizontal) direction. Rectangular
lattices with such short wavelength lattice distortions support an
electron DOS with a gap at the center, if we consider the natural  electron
hopping amplitude  modulation by the changes in interatomic
distances. Therefore, such lattices have an insulating DOS for the
electron density of half an electron per site. Even though the 2D
lattice described above is simple, it shares all the three properties of
manganites that we believe are essential for the observed
inhomogeneity, and provides a testing ground for whether the complex
energy landscape and long range strain-strain interaction can indeed
give rise to self-consistent multiscale inhomogeneities.

\begin{figure}[ht]
\includegraphics[scale=1.0, clip=true]{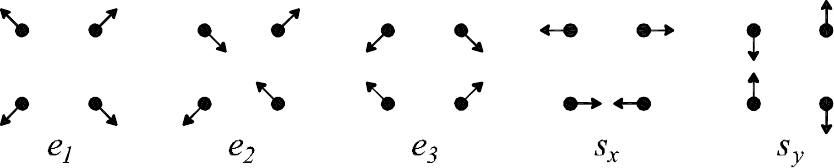}
\caption{Symmetry distortion modes~\cite{Ahn03} for the motif for the two-dimensional square lattice
with a monatomic basis shown in Fig.~\ref{fig:square}.}
\label{fig:modes}
\end{figure}

We require an energy expression for which the undistorted and distorted
lattices described above are the local and global energy minimum
states. For this purpose, we use an atomic-scale mode-based
description of lattice distortions that we developed
recently.~\cite{Ahn03} In this method, we use normal modes of a
square plaquette of four atoms, instead of displacement variables, to
describe lattice distortions. These atomic scale modes for the
monatomic square lattice are shown in Fig.~\ref{fig:modes}. The
first three modes are  long wave length modes, since they can
be obtained by uniformly deforming the square lattice.  The last two
modes, which correspond to $(\pi,\pi)$ staggered distortions of the
lattice, are  short wavelength modes. For a square lattice,
each atom is shared by four neighboring plaquettes, which makes the
modes at neighboring plaquettes dependent on each other. Such
a constraint can be expressed in terms of equations in the Fourier
transformed space, and the five modes can describe any lattice
distortion for the square lattice with a monatomic basis. In the
long wavelength limit, the three long wavelength modes become
identical to the familiar strain modes, which makes our approach ideal
for describing nano-and micro-meter scale inhomogeneities within the same
theoretical framework. The inclusion of constraints allows our method to
automatically generate the effects of the long range anisotropic
interaction, the origin of which is the short-range bonding constraint.

We consider an $N\times N$ square lattice with a modified periodic condition
explained below. The displacement variables for the atom at the site $\vec{i}$
are $u^x_{\vec{i}}$ and $u^y_{\vec{i}}$.
The distance between the nearest neighbor atoms, $a$, is
irrelevant to our formalism presented below,
and can be chosen depending on the relative size of the distortions compared to the lattice constants.
In all the figures in this paper, $a$ is chosen as 10
so that the size of the distortion relative to the interatomic distance
is of the same order of the magnitude as observed in charge and orbital ordered manganites.
In general, the displacement of atoms in a periodic
structure can be described using two components. One is the component
that changes monotonically as the site indices shift along a
direction. This component, represented by a superscript `$nF$'
below, can not be Fourier-transformed and corresponds to
the uniform distortion of the lattice. The rest of the displacement, represented by a superscript
`$F$' below, can be Fourier-transformed and is subject to the
periodic boundary condition. Therefore, we express the displacements
as follows:
\begin{eqnarray}
u^x_{\vec{i}}&=&u^{x,nF}_{\vec{i}}+u^{x,F}_{\vec{i}}, \label{eq:uxi} \\
u^y_{\vec{i}}&=&u^{y,nF}_{\vec{i}}+u^{y,F}_{\vec{i}}, \label{eq:uyi}
\end{eqnarray}
where
\begin{eqnarray}
u^{x,nF}_{\vec{i}}&=& \varepsilon^{xx}_0 i_x + \varepsilon^{xy}_0 i_y, \label{eq:uxnF} \\
u^{y,nF}_{\vec{i}}&=& \varepsilon^{xy}_0 i_x + \varepsilon^{yy}_0 i_y, \label{eq:uynF}
\end{eqnarray}
and
\begin{eqnarray}
u^{x,F}_{\vec{i}}&=&\sum_{\vec{k}} u^x_{\vec{k}}e^{i\vec{k}\cdot\vec{i}}, \\
u^{y,F}_{\vec{i}}&=&\sum_{\vec{k}} u^y_{\vec{k}}e^{i\vec{k}\cdot\vec{i}}.
\end{eqnarray}
We note that $u^x_{\vec{k}}$ and $u^y_{\vec{k}}$ are obtained through the Fourier
transformation of $u^{x,F}_{\vec{i}}$ and $u^{y,F}_{\vec{i}}$,
rather than $u^{x}_{\vec{i}}$ and $u^{y}_{\vec{i}}$. The periodic
boundary condition results in
\begin{eqnarray}
k_x&=&\frac{2\pi n_x}{N}, \\
k_y&=&\frac{2\pi n_y}{N},
\end{eqnarray}
where $n_x=-N/2-1, ... , N/2$ and $n_y=-N/2-1, ... , N/2$. For
$u^{x,nF}_{\vec{i}}$ and $u^{y,nF}_{\vec{i}}$, the rigid rotation of
the whole system is excluded, since it is irrelevant to the
potential energy change. Similarly, the $\vec{k}=0$ components of
$u^{x,F}_{\vec{i}}$ and $u^{y,F}_{\vec{i}}$ correspond to the rigid
translation of the whole system, which are set to zero.

For the square lattice, we define the symmetry modes shown in
Fig.~\ref{fig:modes} as follows:
    \begin{eqnarray}
        e_{1}(\vec{i})&=& \frac{1}{2\sqrt{2}} \left(-u^{x}_{\vec{i}}-u^{y}_{\vec{i}}+u^{x}_{\vec{i}+10}-u^{y}_{\vec{i}+10}
              -u^{x}_{\vec{i}+01}+u^{y}_{\vec{i}+01}+u^{x}_{\vec{i}+11}+u^{y}_{\vec{i}+11}\right), \label{eq:def-e1} \\
        e_{2}(\vec{i})&=& \frac{1}{2\sqrt{2}} \left(-u^{x}_{\vec{i}}-u^{y}_{\vec{i}}-u^{x}_{\vec{i}+10}+u^{y}_{\vec{i}+10}
        +u^{x}_{\vec{i}+01}-u^{y}_{\vec{i}+01}+u^{x}_{\vec{i}+11}+u^{y}_{\vec{i}+11}\right), \\
        e_{3}(\vec{i})&=& \frac{1}{2\sqrt{2}} \left(-u^{x}_{\vec{i}}+u^{y}_{\vec{i}}+u^{x}_{\vec{i}+10}+u^{y}_{\vec{i}+10}
             -u^{x}_{\vec{i}+01}-u^{y}_{\vec{i}+01}+u^{x}_{\vec{i}+11}-u^{y}_{\vec{i}+11}\right), \label{eq:def-e3} \\
        s_{x}(\vec{i})&=& \frac{1}{2} \left(u^{x}_{\vec{i}}-u^{x}_{\vec{i}+10}-u^{x}_{\vec{i}+01}+u^{x}_{\vec{i}+11}\right), \\
        s_{y}(\vec{i})&=& \frac{1}{2} \left(u^{y}_{\vec{i}}-u^{y}_{\vec{i}+10}-u^{y}_{\vec{i}+01}+u^{y}_{\vec{i}+11}\right).
    \end{eqnarray}
These modes are fully subject to the periodic boundary condition,
e.g., $e_1(i_x,i_y)=e_1(i_x+N,i_y)=e_1(i_x,i_y+N)$, unlike
the displacement variables. Thus, they can be Fourier-transformed, for example,
according to
\begin{equation}
e_1(\vec{i})=\sum_{\vec{k}} e_1(\vec{k}) e^{i\vec{k}\cdot\vec{i}}.
\end{equation}
From the definitions, we find
\begin{eqnarray}
e_1(\vec{k}=0)&=&\frac{\varepsilon^{xx}_0+\varepsilon^{yy}_0}{\sqrt{2}} \equiv \tilde{e}_1, \label{eq:e1k0} \\
e_2(\vec{k}=0)&=&\frac{\varepsilon^{xy}_0}{\sqrt{2}} \equiv \tilde{e}_2,                    \label{eq:e2k0} \\
e_3(\vec{k}=0)&=&\frac{\varepsilon^{xx}_0-\varepsilon^{yy}_0}{\sqrt{2}} \equiv \tilde{e}_3, \label{eq:e3k0} \\
s_x(\vec{k}=0)&=&0, \label{eq:sxk0}\\
s_y(\vec{k}=0)&=&0, \label{eq:syk0}
\end{eqnarray}
and
\begin{eqnarray}
        e_{1}(\vec{k}\neq 0) &=& \frac{1}{2\sqrt{2}}\left[-(1-e^{ik_x})(1+e^{ik_y})u^{x}_{\vec{k}}
             -(1+e^{ik_x})(1-e^{ik_y})u^{y}_{\vec{k}}\right], \label{eq:e1kne0}\\
        e_{2}(\vec{k}\neq 0) &=& \frac{1}{2\sqrt{2}}\left[-(1+e^{ik_x})(1-e^{ik_y})u^{x}_{\vec{k}}
             -(1-e^{ik_x})(1+e^{ik_y})u^{y}_{\vec{k}}\right], \label{eq:e2kne0}\\
        e_{3}(\vec{k}\neq 0) &=& \frac{1}{2\sqrt{2}}\left[-(1-e^{ik_x})(1+e^{ik_y})u^{x}_{\vec{k}}
             +(1+e^{ik_x})(1-e^{ik_y})u^{y}_{\vec{k}}\right], \label{eq:e3kne0}\\
        s_{x}(\vec{k}\neq 0) &=&  \frac{1}{2}(1-e^{ik_x})(1-e^{ik_y})u^{x}_{\vec{k}}, \label{eq:sxkne0}\\
        s_{y}(\vec{k}\neq 0) &=&  \frac{1}{2}(1-e^{ik_x})(1-e^{ik_y})u^{y}_{\vec{k}}. \label{eq:sykne0}
\end{eqnarray}
We note that the $\vec{k}=0$ components of the symmetry modes are from
$u^{x,nF}_{\vec{i}}$ and $u^{y,nF}_{\vec{i}}$,
while $\vec{k}=0$ components of $u^{x,F}_{\vec{i}}$ and $u^{y,F}_{\vec{i}}$ do not contribute to the distortion modes.

The five variables are related by three constraint equations,
because only two physically independent displacement
variables exist for each site. As discussed in Ref.~\onlinecite{Ahn03},
these constraint equations are found from the relations between the
symmetry modes and the displacement variables in  reciprocal
space. For $k_x \neq 0$ and $k_y \neq 0$, we invert the linear
relations between [$s_x(\vec{k})$, $s_y(\vec{k})$] and
[$u^x_{\vec{k}}$, $u^y_{\vec{k}}$] in Eqs.~(\ref{eq:sxkne0}) and
(\ref{eq:sykne0}) and replace them in the expressions with other modes in
Eqs.~(\ref{eq:e1kne0})-(\ref{eq:e3kne0}). This  leads to
\begin{eqnarray}
         \sin{\frac{k_x}{2}}\cos{\frac{k_y}{2}}s_{x}(\vec{k}) +
            \cos{\frac{k_x}{2}}\sin{\frac{k_y}{2}}s_{y}(\vec{k})
         - \sqrt{2} i \sin{\frac{k_x}{2}}\sin{\frac{k_y}{2}}e_{1}(\vec{k})
            &=& 0,  \label{eq:constraint1} \\
         \cos{\frac{k_x}{2}}\sin{\frac{k_y}{2}}s_{x}(\vec{k}) +
            \sin{\frac{k_x}{2}}\cos{\frac{k_y}{2}}s_{y}(\vec{k})
         - \sqrt{2} i \sin{\frac{k_x}{2}}\sin{\frac{k_y}{2}}e_{2}(\vec{k})
            &=& 0, \label{eq:constraint2} \\
         \sin{\frac{k_x}{2}}\cos{\frac{k_y}{2}}s_{x}(\vec{k}) -
            \cos{\frac{k_x}{2}}\sin{\frac{k_y}{2}}s_{y}(\vec{k})
         - \sqrt{2} i \sin{\frac{k_x}{2}}\sin{\frac{k_y}{2}}e_{3}(\vec{k})
            &=& 0. \label{eq:constraint3}
\end{eqnarray}
These constraint equations indicate
that $e_1(\pi,\pi)$, $e_2(\pi,\pi)$, and $e_3(\pi,\pi)$ vanish and $\tilde{s}_x \equiv
s_x(\pi,\pi)$ and $\tilde{s}_y \equiv s_y(\pi,\pi)$ are independent variables.
Constraint equations for $k_x=0$ or $k_y=0$ should be
considered separately from
Eqs.~(\ref{eq:constraint1})-(\ref{eq:constraint3}).
Equations~(\ref{eq:e1k0})-(\ref{eq:syk0}) show that
$e_1(\vec{k}=0)$, $e_2(\vec{k}=0)$ and $e_3(\vec{k}=0)$ are
independent of each other, and $s_x(\vec{k}=0)$ and $s_y(\vec{k}=0)$
vanish. For $k_x=0$ and $k_y \neq 0$,
Eqs.~(\ref{eq:e1kne0})-(\ref{eq:sykne0}) show that
$e_1(\vec{k})=-e_3(\vec{k})$ and $e_2(\vec{k})$ are independent variables, and
$s_x(\vec{k})=s_y(\vec{k})=0$. Similarly, for $k_x \neq
0$ and $k_y = 0$, $e_1(\vec{k})=e_3(\vec{k})$ and $e_2(\vec{k})$
are independent variables, and $s_x(\vec{k})=s_y(\vec{k})=0$.

To describe lattice distortions in our simulations,
we primarily use the variables $s_x(\vec{i})$ and $s_y(\vec{i})$.
These variables
can be assigned arbitrarily except that they should satisfy
$s_x(\vec{k})=s_y(\vec{k})=0$ if $k_x=0$ or $k_y=0$, as
required by Eqs.~(\ref{eq:sxk0}), (\ref{eq:syk0}),
(\ref{eq:sxkne0}), and (\ref{eq:sykne0}). In our numerical
simulations, we implement this condition by subtracting
unphysical components with $k_x=0$ or $k_y=0$ from $s_x(\vec{i})$
and $s_y(\vec{i})$, each time we initialize or change $s_x(\vec{i})$
and $s_y(\vec{i})$.
However, $s_x(\vec{i})$ and $s_y(\vec{i})$ do not uniquely determine lattice distortions,
because of the singular relation between [$s_x(\vec{k})$, $s_y(\vec{k})$]
and [$u^x_{\vec{k}}$, $u^y_{\vec{k}}$] in Eqs.~(\ref{eq:sxkne0}) and (\ref{eq:sykne0}).
As seen above, $e_1(\vec{k}=0)$, $e_2(\vec{k}=0)$, $e_3(\vec{k}=0)$,
$e_1(k_x=0, k_y \neq 0)=-e_3(k_x=0, k_y \neq 0)$, $e_2(k_x=0, k_y \neq 0)$,
$e_1(k_x \neq 0, k_y = 0)= e_3(k_x \neq 0, k_y = 0)$, and $e_2(k_x \neq 0, k_y = 0)$
should be specified, in addition to $s_x(\vec{i})$ and $s_y(\vec{i})$,
for the complete description of lattice distortions.

From these variables, displacement variables,
$u^x_{\vec{i}}$ and $u^y_{\vec{i}}$, are calculated.
For the non-periodic parts of displacements, $u_{\vec{i}}^{x,nF}$ and
$u_{\vec{i}}^{y,nF}$ in Eqs.~(\ref{eq:uxnF}) and (\ref{eq:uynF}), we
use Eqs.~(\ref{eq:e1k0})-(\ref{eq:e3k0}) to obtain
\begin{eqnarray}
u_{\vec{i}}^{x,nF}&=&\frac{e_1(\vec{k}=0)+e_3(\vec{k}=0)}{\sqrt{2}}i_x+\sqrt{2}e_2(\vec{k}=0)i_y, \label{eq:uxnFe123} \\
u_{\vec{i}}^{y,nF}&=&\sqrt{2}e_2(\vec{k}=0) i_x +\frac{e_1(\vec{k}=0)-e_3(\vec{k}=0)}{\sqrt{2}}i_y. \label{eq:uynFe123}
\end{eqnarray}
We find the periodic parts of the displacement, $u_{\vec{i}}^{x,F}$
and $u_{\vec{i}}^{y,F}$, through the Fourier transformation of
$u^x_{\vec{k}}$ and $u^y_{\vec{k}}$, which are obtained by inverting
two non-singular equations among
Eqs.~(\ref{eq:e1kne0})-(\ref{eq:sykne0}).
Therefore, if $k_x\neq0$ and $k_y\neq0$, we invert Eqs.~(\ref{eq:sxkne0}) and
(\ref{eq:sykne0}) to obtain
\begin{eqnarray}
&& u^x_{k_x\neq 0,k_y \neq 0}=\frac{2}{(1-e^{ik_x})(1-e^{ik_y})} s_x(\vec{k}),  \label{eq:uxxn0yn0} \\
&& u^y_{k_x\neq 0,k_y \neq 0}=\frac{2}{(1-e^{ik_x})(1-e^{ik_y})} s_y(\vec{k}). \label{eq:uyxn0yn0}
\end{eqnarray}
If $k_x\neq0$ and $k_y=0$, Eqs.~(\ref{eq:e1kne0}) and (\ref{eq:e2kne0}) lead to
\begin{eqnarray}
&& u^x_{k_x\neq 0, k_y=0}= -\frac{\sqrt{2}} {1-e^{ik_x}} e_1(\vec{k}), \\
&& u^y_{k_x\neq 0, k_y=0}= -\frac{\sqrt{2}} {1-e^{ik_x}} e_2(\vec{k}).
\end{eqnarray}
Similarly, if $k_x=0$ and $k_y\neq0$, we obtain
\begin{eqnarray}
&& u^x_{k_x= 0, k_y \neq 0} = \frac{\sqrt{2}}{1-e^{ik_y}} e_2(\vec{k}), \\
&& u^y_{k_x= 0, k_y \neq 0} = \frac{\sqrt{2}}{1-e^{ik_y}} e_1(\vec{k}).
\end{eqnarray}
The  $k_x=0$ and $k_y=0$ components of the displacements correspond
to rigid displacements, which are set to zero:
\begin{eqnarray}
&& u^x_{k_x= 0, k_y = 0}= 0, \label{eq:uxxe0ye0} \\
&& u^y_{k_x= 0, k_y = 0}= 0. \label{eq:uyxe0ye0}
\end{eqnarray}
By adding periodic and non-periodic parts of displacements according to
Eqs.~(\ref{eq:uxi}) and (\ref{eq:uyi}), we find
$u^x_{\vec{i}}$ and $u^y_{\vec{i}}$.

\subsection{Total energy of the model and the Hamiltonian for electronic property calculations}
In terms of the above modes, we consider the following energy
expression, $E_{tot}$, as the total energy of a model system for
strain-induced phase coexistence:~\cite{Ahn04}
\begin{eqnarray}
&&E_{tot}=E_{s}+E_{l}+E_{c}, \label{eq:Etot} \\
&&E_{s}=\sum_{\vec{i}} \left[ \frac{B}{2} (s_x^2+s_y^2) + \frac{G_1}{4}
(s_x^4 + s_y^4) + \frac{G_2}{2} s_x^2 s_y^2
 + \frac{H_1}{6} (s_x^6 + s_y^6) +\frac{H_2}{6} s_x^2 s_y^2
(s_x^2+s_y^2)\right]_{\vec{i}}, \label{eq:Es} \\
&&E_{l}= \sum_{\vec{i}} \left[ \frac{A_1}{2} e_1^2 + \frac{A_2}{2} e_2^2
+ \frac{A_3}{2} e_3^2 \right]_{\vec{i}}, \label{eq:El} \\
&&E_{c}= \sum_{\vec{i}} \left[ C_3 (s_x^2-s_y^2) e_3 \right]_{\vec{i}}. \label{eq:Ec}
\end{eqnarray}
The first term $E_s$ with short wavelength modes includes all
symmetry-allowed terms up to  sixth order with all coefficients
positive, since we are interested in a first-order-transition-like energy
landscape. The second term $E_l$ with long wavelength modes up to
 second order mediates the long range anisotropic interactions.
The third term $E_{c}$ represents the coupling between the long and
the short wavelength modes, where the $e_3$ mode is coupled to the $s_x$
and $s_y$ modes in a symmetry-allowed form.  This last term gives
rise to the global energy minimum state with long and short
wavelength distortions, in addition to the local energy minimum
state without distortion. The energy expression $E_{tot}$ gives rise
to the desired energy landscape in appropriate ranges of parameters.

To establish a connection between the electronic properties and the
lattice distortions as observed in manganites, namely metallic and
insulating states for the undistorted and distorted phases,
respectively, we use the following Su-Shrieffer-Heeger (SSH)
Hamiltonian for the electronic structure calculations in our
model:
\begin{eqnarray}
H_{SSH}&=&\sum_{\vec{i}}-t_0\left[1-\alpha(u^x_{\vec{i}+(10)}-u^x_{\vec{i}})\right]
\left(c^{\dagger}_{\vec{i}}c_{\vec{i}+(10)}+c^{\dagger}_{\vec{i}+(10)}c_{\vec{i}}\right) \nonumber \\
& &-t_0\left[1-\alpha(u^y_{\vec{i}+(01)}-u^y_{\vec{i}})\right]
\left(c^{\dagger}_{\vec{i}}c_{\vec{i}+(01)}+c^{\dagger}_{\vec{i}+(01)}c_{\vec{i}}\right). \label{eq:SSH}
\end{eqnarray}
Here, we consider only one orbital per site and neglect the electron
spin for simplicity. The operator $c^{\dagger}_{\vec{i}}$ is the creation operator
of an electron at a site $\vec{i}$. In this Hamiltonian,
the electron hopping amplitude is assumed to be linearly modified by the change in the
nearest neighbor interatomic distances. We make the  adiabatic assumption  that the
total energy $E_{tot}$ is obtained by minimizing the energy of the
system with respect to all degrees of freedom, including the
electronic one, except for the lattice degrees of freedom.
Therefore, $E_{tot}$ is used for the calculation of the energy landscape
and the Euler simulations, whereas $H_{SSH}$ is used for the calculations of
electronic properties associated with templates of lattice
distortions.

The SSH Hamiltonian is a Hamiltonian for independent electrons, and
can be diagonalized within a one-electron basis. Therefore, we
construct electronic Hamiltonian matrices for given lattice
distortions, $u^x_{\vec{i}}$ and $u^y_{\vec{i}}$, with the basis set
 $\{ c^{\dagger}_{\vec{i}} | 0 \rangle \}$, where $| 0 \rangle$
represents the state without electrons.  We diagonalize the matrices
numerically, and fill the eigenstates with electrons according to the electron
density. Representing the $l$-th lowest energy eigenstate as
\begin{equation}
| l \rangle = \sum_{\vec{i}} z_{l,\vec{i}} c^{\dagger}_{\vec{i}} | 0 \rangle,
\end{equation}
the local DOS at site $\vec{i}$ is calculated by
\begin{equation}
D_{\vec{i}}(E)=\sum_l \delta (E-E_l) |z_{l,\vec{i}}|^2 ,
\end{equation}
which reveals local electronic properties. The same approach has
been used in Ref.~\onlinecite{Ahn05} to study electronic
inhomogeneities around structural twin and antiphase boundaries in
model systems with a strong electron-lattice coupling.

\subsection{Energy landscape for homogeneous states}
We expect the ground state of the model energy expression $E_{tot}$
to be homogeneous with $\tilde{e}_1$, $\tilde{e}_2$,
$\tilde{e}_3$, $\tilde{s}_x$ and $\tilde{s}_y$ distortions only, defined
in Eqs.~(\ref{eq:e1k0})-(\ref{eq:e3k0}) and below Eq.~(\ref{eq:constraint3}),
considering the way that the energy terms are selected. Therefore, we study
the following energy expression, which includes these particular distortions only,
to understand the energy landscape for the homogeneous states:
\begin{eqnarray}
E_{tot}^h&=&E_{s}^h+E_{l}^h+E_{c}^h, \\
\frac{E_{s}^h}{N^2}&=& \frac{B}{2} \left(\tilde{s}_x^2+\tilde{s}_y^2\right) + \frac{G_1}{4}
\left(\tilde{s}_x^4 + \tilde{s}_y^4\right) + \frac{G_2}{2} \tilde{s}_x^2 \tilde{s}_y^2
+ \frac{H_1}{6} \left(\tilde{s}_x^6 + \tilde{s}_y^6\right) +\frac{H_2}{6} \tilde{s}_x^2 \tilde{s}_y^2
\left(\tilde{s}_x^2+\tilde{s}_y^2\right), \\
\frac{E_{l}^h}{N^2}&=& \frac{A_1}{2} \tilde{e}_1^2 + \frac{A_2}{2} \tilde{e}_2^2
+ \frac{A_3}{2} \tilde{e}_3^2, \\
\frac{E_{c}^h}{N^2}&=& C_3 \left(\tilde{s}_x^2-\tilde{s}_y^2\right) \tilde{e}_3.
\end{eqnarray}
Because $\tilde{e}_1$, $\tilde{e}_2$, $\tilde{e}_3$, $\tilde{s}_x$
and $\tilde{s}_y$ are independent of each other, we minimize
$E_{tot}^h$ with respect to $\tilde{e}_1$, $\tilde{e}_2$, and
$\tilde{e}_3$ independently and obtain $\tilde{e}_1=0$,
$\tilde{e}_2=0$, and
$\tilde{e}_3=-C_3(\tilde{s}_x^2-\tilde{s}_y^2)/A_3$. We insert these
back into $E_{tot}^h$ and obtain the following energy expression
 in terms of $\tilde{s}_x$ and $\tilde{s}_y$ only:
\begin{eqnarray}
\frac{E_{tot}^{h,min}}{N^2}&=&\frac{B}{2}\left(\tilde{s}_{x}^2+\tilde{s}_{y}^2\right)
+\frac{1}{4}\left(G_1-2\frac{C_3^2}{A_3}\right)\left(\tilde{s}_x^4+\tilde{s}_y^4\right)
+\frac{1}{2}\left(G_2+2\frac{C_3^2}{A_3}\right)\tilde{s}_x^2 \tilde{s}_y^2  \nonumber \\
&&+\frac{H_1}{6} \left(\tilde{s}_x^6 + \tilde{s}_y^6\right) +\frac{H_2}{6}
\tilde{s}_x^2 \tilde{s}_y^2 \left(\tilde{s}_x^2+\tilde{s}_y^2\right).
\label{eq:Ehmintot}
\end{eqnarray}
We find parameter values, for which $E_{tot}^{h,min}/N^2$ has one
local energy minimum state without distortion and four
symmetry-related degenerate global energy minimum states with
distortions. Necessary conditions for such first-order-transition-like energy
landscape are $G_1' \equiv G_1-2 C_3^2/A_3 < 0$ and $(G_1')^2
> 4 B H_1 $, for which the energy minima occur at
$(\tilde{s}_x, \tilde{s}_y) = (0,0)$,
$(\pm s_0,0)$, and $(0, \pm s_0)$ with
\begin{equation}
s_0=\sqrt{\frac{-G_1'+\sqrt{(G_1')^2-4BH_1}}{2 H_1}}.
\end{equation}
The locations of the five energy minima in the $\tilde{s}_x-\tilde{s}_y$
plane are indicated in Fig.~\ref{fig:energy}(a). Corresponding
uniform mode distortions are $\tilde{e_3}=-C_3 s_0^2/A_3$ for
$(\tilde{s}_x,\tilde{s}_y)=(\pm s_0,0)$, and $\tilde{e_3}=C_3
s_0^2/A_3$ for $(\tilde{s}_x,\tilde{s}_y)=(0,\pm s_0)$. For
comparison, we choose two sets of parameter values, one giving a
shallow and the other a deep local energy minimum at
$(\tilde{s}_x, \tilde{s}_y) = (0,0)$, as shown with a thin blue and a thick red curve,
respectively, in Fig.~\ref{fig:energy}(b).
In manganites, the
difference in the depth of the energy landscape can be related to
the size of rare earth or alkali metal elements, which is known experimentally to
influence the physical properties of manganites.~\cite{Hwang95}
Alternatively, we may consider this a measure of ``microstrain".~\cite{Bianconi}
\subsection{Methods of simulations for inhomogeneous states}
The energy landscape is much more complicated for inhomogeneous
states because of the constraint relations among the distortion modes.
To study inhomogeneous configurations, particularly, metastable configurations, we first minimize
$E_{tot}$ analytically with respect to all the independent variables except $s_x(\vec{i})$ and $s_y(\vec{i})$, that is,
$e_1(\vec{k}=0)$, $e_2(\vec{k}=0)$, $e_3(\vec{k}=0)$,
$e_1(k_x=0, k_y \neq 0)=-e_3(k_x=0, k_y \neq 0)$, $e_2(k_x=0, k_y \neq 0)$,
$e_1(k_x \neq 0, k_y = 0)= e_3(k_x \neq 0, k_y = 0)$, and $e_2(k_x \neq 0, k_y = 0)$,
and obtain an energy expression $E'_{tot}(s_x,s_y)$.
The details of the derivation and expression for $E'_{tot}(s_x,s_y)$ are provided in Appendix A.

In our simulations, we set initial configurations of $s_x(\vec{i})$
and $s_y(\vec{i})$ and relax the lattice according to the Euler method,
\begin{eqnarray}
s_x^{n+1}(\vec{i})&=&s_x^{n}(\vec{i})-\gamma \left.\frac{\partial E'_{tot}(s_x,s_y)}{\partial s_x(\vec{i})}\right|_{s_x^{n},s_y^{n}}, \\
s_y^{n+1}(\vec{i})&=&s_y^{n}(\vec{i})-\gamma \left.\frac{\partial E'_{tot}(s_x,s_y)}{\partial s_y(\vec{i})}\right|_{s_x^{n},s_y^{n}},
\end{eqnarray}
where the superscript $n$ or $n+1$ represents the number of  Euler
steps taken from the initial configuration, and $\gamma$ controls
the size of the Euler step.
Expressions for $\partial E'_{tot}(s_x,s_y) / \partial s_x(\vec{i})$ and
$\partial E'_{tot}(s_x,s_y)/\partial s_y(\vec{i})$ are provided in Appendix B.
We change $s_x(\vec{i})$ and
$s_y(\vec{i})$ for all $\vec{i}$'s simultaneously at each step.
We run the simulation until $E'_{tot}(s_x,s_y)$ does not decrease
 further, but only fluctuates, which is an indication that the system
has reached a local energy minimum configuration.

\subsection{Initial conditions and results of the simulations for inhomogeneous states}
\begin{figure}[ht]
\includegraphics[scale=0.5, clip=true]{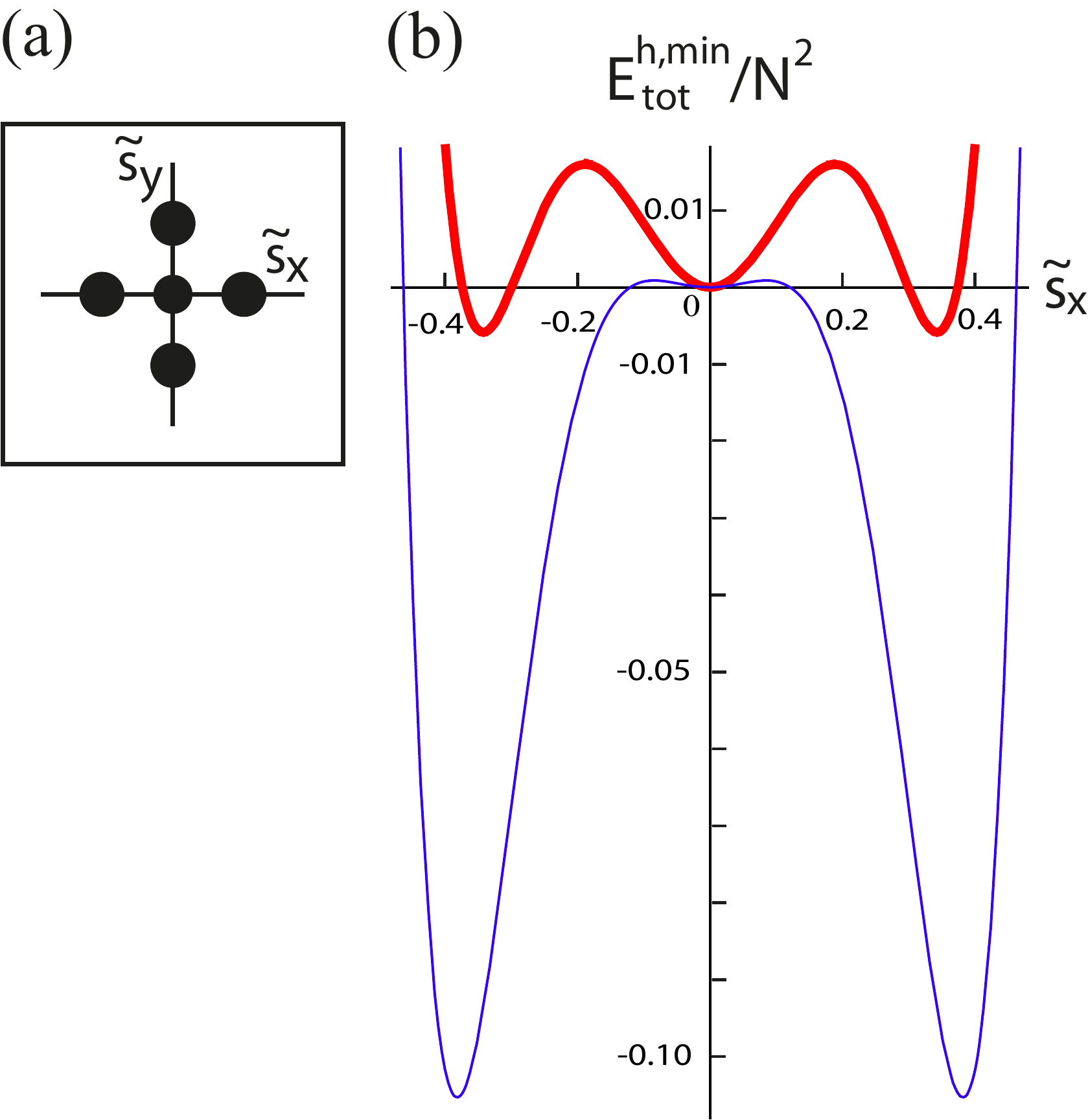}
\caption{\label{fig:energy} (Color online) Energy landscape for the homogeneous phase.
(a) Solid
dots mark the locations of local and global energy minima
for $E^{h,min}_{tot}/N^2$ in Eq.~(\ref{eq:Ehmintot})
in the $\tilde{s}_x-\tilde{s}_y$ plane,
where $\tilde{s}_x=s_x(\pi,\pi)$
and $\tilde{s}_y=s_y(\pi,\pi)$.
The local energy minimum states with the opposite signs of $\tilde{s}_x$ or $\tilde{s}_y$
are related by a phase difference of the short wavelength distortion,
whereas the local energy minimum states along $\tilde{s}_x$ axis
is related to those along $\tilde{s}_y$ axis by the exchange of $x$ and $y$ axes.
(b) $E^{h,min}_{tot}/N^2$ versus
$\tilde{s}_x$ with $\tilde{s}_y=0$ for the two chosen parameter
sets. The parameter values for
the deep local energy minimum case, represented by the thick red curve, are $A_1=7$, $A_2=4$, $A_3=6$,
$B=2$, $C_3=20$, $G_1=60$, $G_2=80$, $H_1=480$, and $H_2=640$, for
which $G'_1=-73.3$, $G'_2 \equiv G_2+2C_3^2/A_3=213$, $s_0=0.34$,
and $|\tilde{e}_3|=0.39$, and $E_{tot}^{h,min}/N^2$ has a value of
-0.0058 for the global energy minimum states.
The height of the energy barrier between local and global energy minima is
0.0160, measured from the local energy minimum. The only different
parameter value for the shallow local minimum case, represented by the thin blue curve, is $B=0.5$, which
gives rise to $s_0=0.38$, $|\tilde{e}_3|=0.49$, the global energy
minimum $E_{tot}^{h,min}/N^2$ of -0.1053, and energy barrier of 0.0009.
Such difference in the energy landscape can be related, for example, to the average size of the rare earth
and alkali metal elements in manganites.}
\end{figure}

\begin{figure}[ht]
\epsfxsize8.0cm\epsfbox{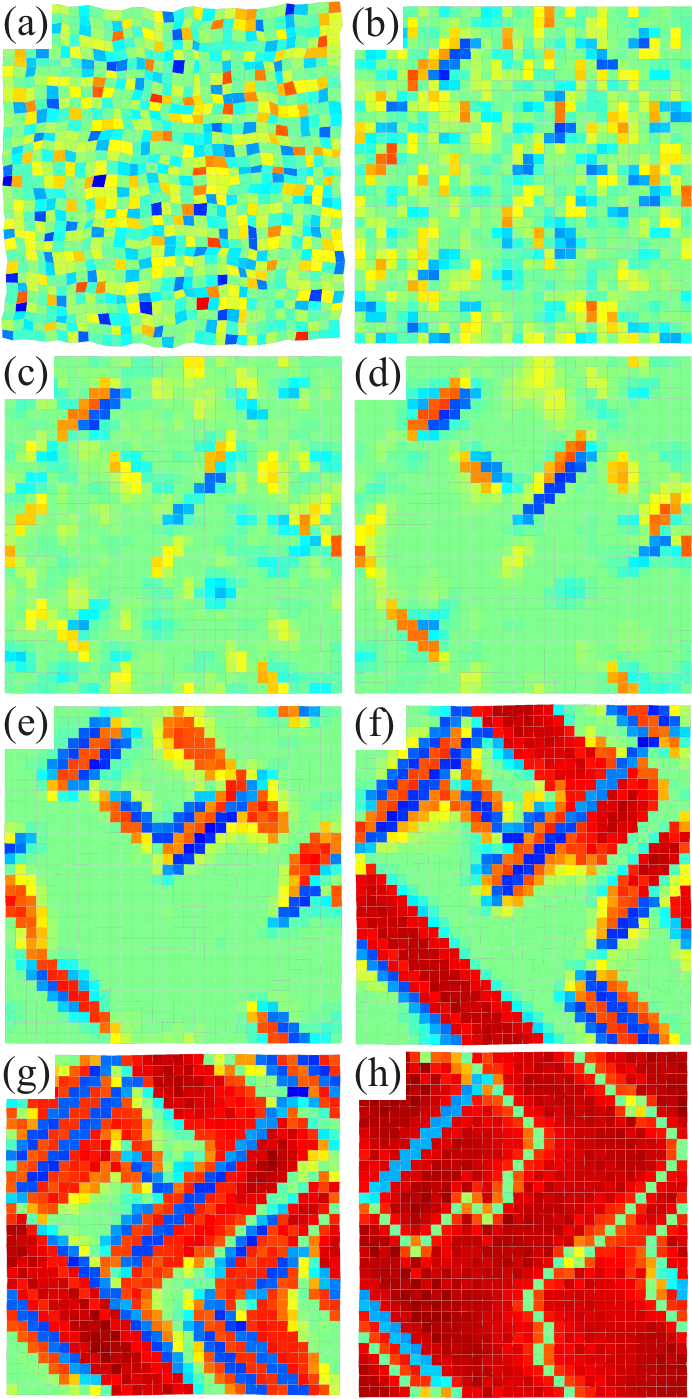} \caption{\label{fig:shallow1}
(Color online) Simulation of lattice relaxation for a $32 \times 32$ lattice for the energy landscape with a
shallow local minimum, namely the thin blue curve in
Fig.~\ref{fig:energy}(b).
The vertices correspond to the
positions of atoms and actual distortions are shown.
The color represents $p_3=s_x^2-s_y^2$,
green corresponding to zero, red and blue $\pm s_0^2$, except
$\pm (2.6 s_0)^2$ for the panel (a).
The dynamics is governed by the Euler
method, simulating a rapid quenching
from a random initial configuration shown in (a).
(b)-(g) show intermediate configurations and (h) the final stable
configuration, in which most region is changed to distorted state
except the antiphase boundaries represented in green.
Highly anisotropic nucleation and the effect of long range interaction
can be identified, for example, in (d) and (e).
}
\end{figure}

We describe initial conditions, parameters, and results of
the simulations in this subsection. Figure~\ref{fig:shallow1} shows the
results of the simulations carried out on a $32\times32$ lattice for
the energy landscape with a shallow local minimum, shown in thin blue
curve in Fig.~\ref{fig:energy}(b) for homogeneous states. The color of each plaquette represents
$p_3(\vec{i}) \equiv s_x(\vec{i})^2-s_y(\vec{i})^2$, and
the vertices and distortions of the plaquettes represent the
actual locations of atoms and actual distortions.  Through the coupling
between $p_3$ and $e_3$ in $E_c$, positive and negative values of $p_3$ are
usually accompanied by an $e_3$ distortion elongated along $y$ and $x$
direction, respectively. Most plaquettes with $p_3$ close to zero
have little distortion. Starting from an initial configuration of
$s_x(\vec{i})$ and $s_y(\vec{i})$, randomly chosen between $-2s_0$
and $2s_0$, as shown in Fig.~\ref{fig:shallow1}(a), the system is
relaxed through the Euler method with $\gamma=0.0015$.
Figures~\ref{fig:shallow1}(b)-\ref{fig:shallow1}(h) correspond to the configurations at
the Euler step $n=$100, 400, 1000, 2000, 4000, and 6000, and the stable
configuration at $n=$100000, respectively.~\cite{Final}

Following the energy gradient from a random initial configuration,
the simulation approximately represents a rapid quenching
of the system from a very high temperature to 0 K. The result shows
that most of the system initially changes into the undistorted state, as
suggested in Fig.~\ref{fig:shallow1}(d). Random fluctuations of
$p_3$ tend to cancel each other when averaged over several
interatomic distances, which prevents most regions from evolving into
a global energy minimum state with lattice distortions. Some regions
with relatively large distortions in the initial configuration evolve
into a distorted state, as shown in Figs.~\ref{fig:shallow1}(b)-\ref{fig:shallow1}(d),
and nucleate the distorted phase. These
distorted regions expand into the undistorted region
[Figs.~\ref{fig:shallow1}(e)-\ref{fig:shallow1}(g)], and eventually
transforms most of the system into a distorted global energy minimum state
separated by anti-phase boundaries, as shown in Fig.~\ref{fig:shallow1}(h).
We also find  that there is a critical size for this nucleation: a
distorted region of a small size in Fig.~\ref{fig:shallow1}(c) changes
into the undistorted state in Fig.~\ref{fig:shallow1}(d). Such
nucleation and growth observed in our simulations of the rapid quenching, and
the presence of the critical size for the nucleation, reflect the
first-order-transition-like energy landscape, and are features observed even in
systems with only short range interactions.

\begin{figure}[ht]
\leavevmode \epsfxsize8.5cm\epsfbox{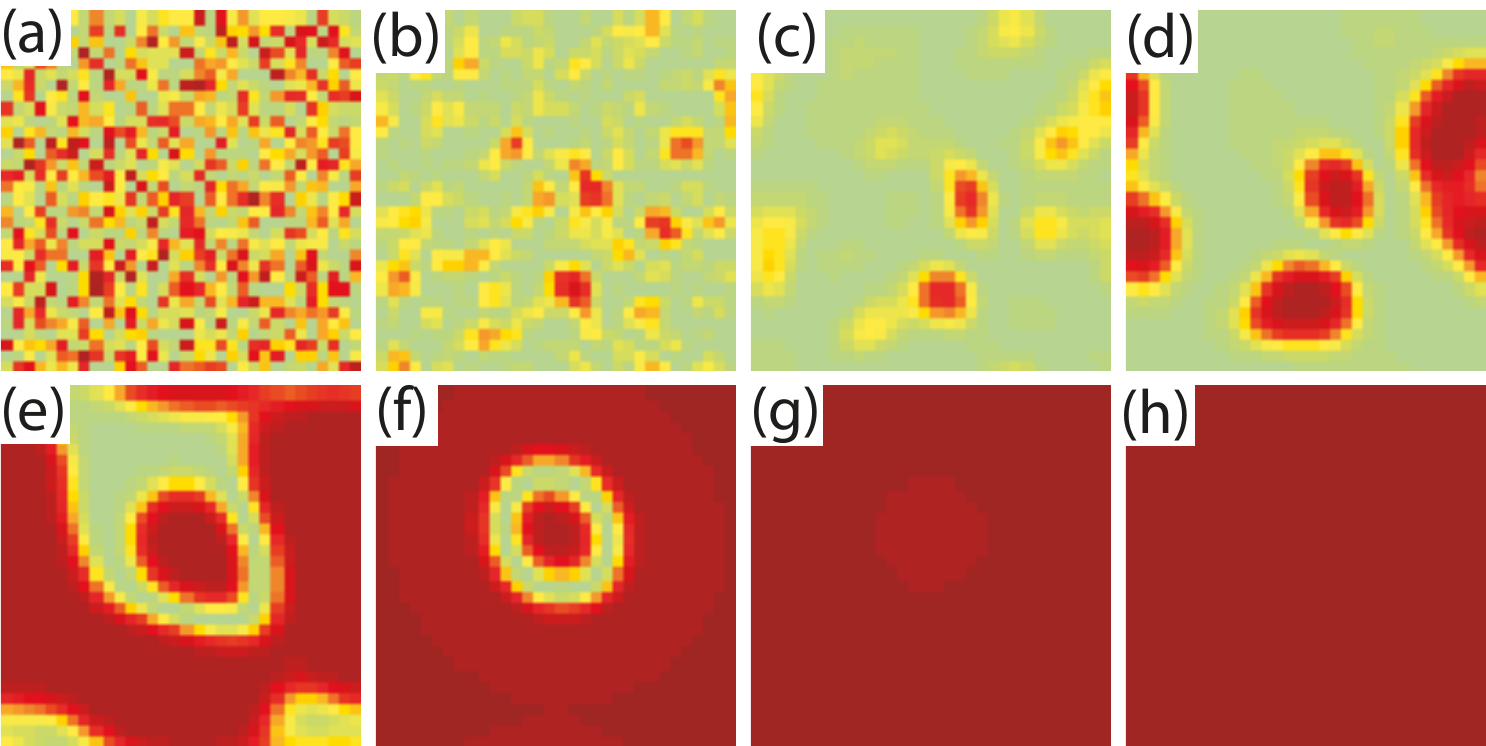}
\caption{\label{fig:compare} (Color online)
Simulation of relaxation for a system
with a short range interaction only.  The variable $p=M^2$ is
plotted on a $32 \times 32$ lattice with a periodic boundary
condition.
Red represents $p= 28.6 M_0^2$ for (a) and
$p=M_0^2$ for (b)-(h), and green $p=0$.
A random initial configuration is shown in (a). (b)-(g)
show intermediate configurations, and (h) the final stable
configuration.
The result shows that nucleation for short range interaction
is nearly isotropic, unlike the case in Fig.~\ref{fig:shallow1}.
}
\end{figure}

For contrast, we study a system with a similar energy landscape,
but without the long range interaction. For this we consider the following
energy expression:
\begin{eqnarray}
E_M&=&\sum_{\vec{i}} \frac{W}{2} M(\vec{i})^2 + \frac{X}{4} M(\vec{i})^2 + \frac{Y}{6} M(\vec{i})^6  \nonumber \\
& &+\frac{Z}{2} \left\{ [M(\vec{i}+(1,0))-M(\vec{i})]^2 + [M(\vec{i}+(0,1))-M(\vec{i})]^2 \right\}, \label{eq:EM}
\end{eqnarray}
where $M(\vec{i})$ is defined at each site.
The last term becomes $Z (\nabla M)^2 /2$, the familiar Ginzburg-Landau gradient term,
in the continuum limit.
The gradient term of $E_M$ with respect to $M(\vec{i})$ for the Euler method is
\begin{eqnarray}
\frac{\partial E_M}{\partial M(\vec{i})} &=& W M(\vec{i}) + X M(\vec{i})^3  + Y M(\vec{i})^5 \nonumber \\
&+&Z [ 4 M(\vec{i}) - M(\vec{i}-(1,0)) - M(\vec{i}+(1,0))
  - M(\vec{i}-(0,1)) - M(\vec{i}+(0,1)) ].
\end{eqnarray}
The parameters $W$, $X$, and $Y$ are chosen to be identical to the
parameters $B$, $G'_1$, and $H_1$ for the lattice model with a
shallow local minimum in Fig.~\ref{fig:energy}(b). We choose $Z=5$,
similar to $A_1$, $A_2$ and $A_3$, since $e_1(\vec{i})$,
$e_2(\vec{i})$, and $e_3(\vec{i})$ are related to the gradient of
$s_x(\vec{i})$ and $s_y(\vec{i})$ in the continuum
limit.~\cite{Ahn03} The uniform ground state for $E_M$ is
$M(\vec{i})=\pm M_0$, where
\begin{equation}
M_0=\sqrt{\frac{-X+\sqrt{X^2-4WY}}{2 Y}}.
\end{equation}
The selected parameter values result in $M_0=0.38$, identical to  $s_0$
for the lattice distortion model.
Figure~\ref{fig:compare} shows the
results of the simulations on a $32\times32$ lattice, in which
$p=M^2$, analogous to $p_3$ for the lattice distortion model, is
shown. The initial
$M(\vec{i})$, shown in Fig.~\ref{fig:compare}(a), is chosen randomly
between $-5.3 M_0$ and $5.3 M_0$, and is relaxed according to the
Euler method with $\gamma=0.00015$. The configurations at the Euler step
$n=$ 1000, 4000, 10000, 2000, 40000, and 60000 are shown in
Figs.~\ref{fig:compare}(b)-\ref{fig:compare}(g). The final stable configuration at
$n=$ 1000000 is shown in Fig.~\ref{fig:compare}(h). These
Euler steps are chosen so that they are consistent with the Euler
steps in Fig.~\ref{fig:shallow1} after being multiplied by
$\gamma$. The system with a short range anisotropic interaction
in Fig.~\ref{fig:compare} also shows nucleation and growth of the low energy phase. However,
comparing Figs.~\ref{fig:shallow1} and ~\ref{fig:compare}
reveals distinct features present only in the lattice distortion model.

First, the nucleation droplets in
Figs.~\ref{fig:shallow1}(c)-\ref{fig:shallow1}(e) are highly
anisotropic, in contrast with those in
Figs.~\ref{fig:compare}(c)-\ref{fig:compare}(e). Second, distortions
separated by relatively large distance along the diagonal direction
interact with each other, grow toward each other, and merge through
the long range interaction, as seen for the yellow and red band along
the 135 degree orientations in
Figs.~\ref{fig:shallow1}(c)-\ref{fig:shallow1}(f). Third, the
nucleation occurs via pairs of distortions with different
orientations to minimize the interface energy between the distorted
region and the undistorted background, as shown in
Figs.~\ref{fig:shallow1}(c) and \ref{fig:shallow1}(d). Such features
are absent in Fig.~\ref{fig:compare}, where the interaction is purely
short-ranged. Recent x-ray scattering experiments have revealed the
presence of short-range anisotropic precursor correlations in
the orthorhombic phase of manganites at high temperatures, which
disappear in the rhombohedral phase.~\cite{Kiryukhin04} Such a feature has
a similarity with the anisotropic droplets observed in our simulations,
and is reminiscent of the precursor embryonic fluctuation
in martensitic transformations.~\cite{Seto90} The quasi-elastic central
peak observed in manganites~\cite{Lynn96,Lynn97} near the metal-insulator
phase transition temperature is also likely to have a structural
origin, similar to the central peak observed in ferroelectrics. Such
experimental observations~\cite{Kim00} and similar features seen in our
simulations indicate that the strain plays an important role in the
formation of nanometer scale inhomogeneity in manganites.

\begin{figure}[ht]
\leavevmode \epsfxsize8.5cm\epsfbox{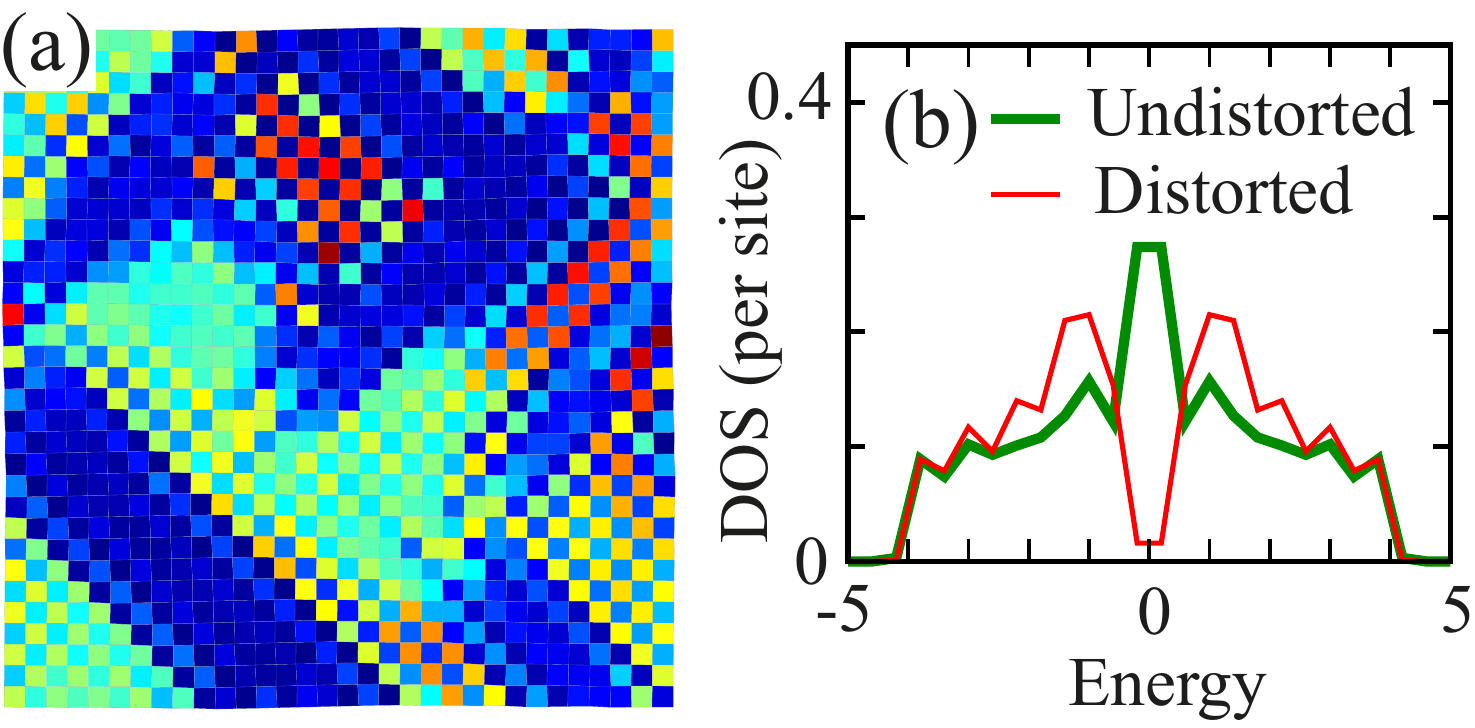}
\caption{\label{fig:ldos_shallow} (Color online)
(a) Map of local electron DOS at $E=0$
calculated for the distortion shown in Fig.~\ref{fig:shallow1}(f).
Blue, green, and red correspond to zero, 0.25, and 0.5 state
per site per unit energy, respectively. (b) Local DOS calculated at
$\vec{i}=(17,10)$, the center of an undistorted metallic region, and
at $\vec{i}=(7,9)$, center of a distorted insulating region.
It shows coexistence of metallic and insulating regions.
}
\end{figure}

To demonstrate electronic inhomogeneity associated with the
structural inhomogeneity, we calculate electronic properties for the
template of the lattice distortions in Fig.~\ref{fig:shallow1}(f).
We use the SSH Hamiltonian in Eq.~(\ref{eq:SSH}) with $t_0=1$ and
$\alpha=1$. The typical local electron densities of states within
undistorted and distorted regions are shown in
Fig.~\ref{fig:ldos_shallow}(b). The local DOS is symmetric about
$E=0$, and a gap (or a ``pseudogap") opens near $E=0$ in distorted regions.
The small DOS within the gap for the distorted region is due to
electron wavefunctions exponentially decaying from the undistorted region.
Therefore,
distorted and undistorted regions have insulating and metallic
electron DOS at a half filling without any spatial charge
inhomogeneity. The map of the local electron DOS at $E=0$ is shown
in Fig.~\ref{fig:ldos_shallow}(a).
The possible inhomogeneity in local DOS without any charge inhomogeneity
in our model is in contrast with other explanations for the inhomogeneity
based on electronic phase separation, an idea similar to the phase
separation in binary alloys.~\cite{Moreo99}

\begin{figure}[ht]
\leavevmode \epsfxsize8.5cm\epsfbox{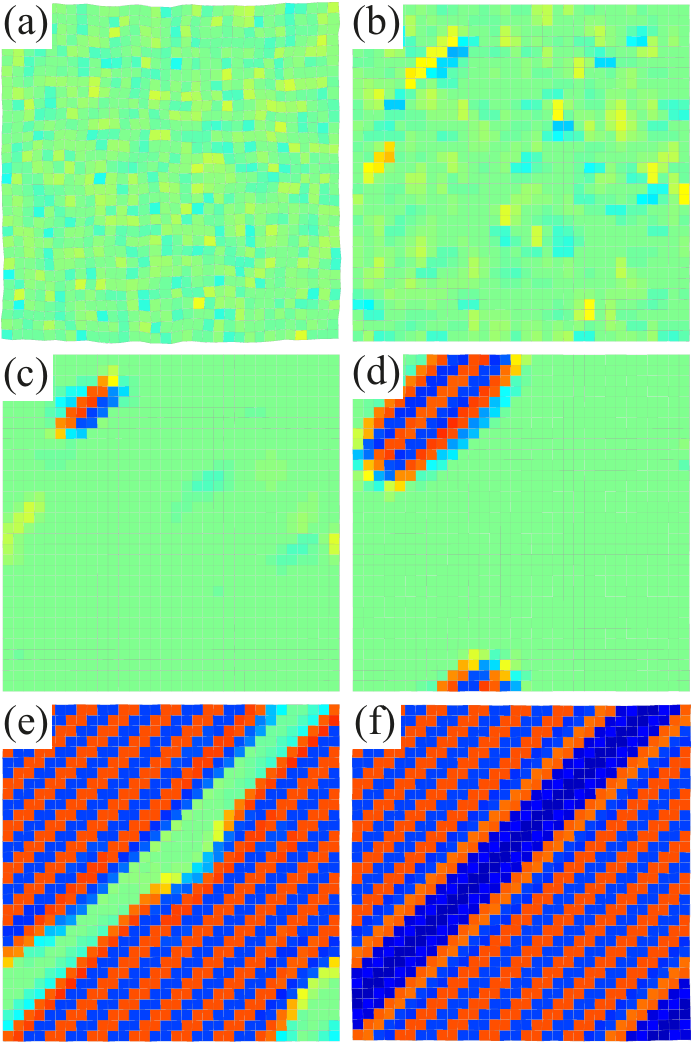}
\caption{\label{fig:shallow2} (Color online)
Simulation of relaxation of lattice
distortions for random initial distortions in a narrower range
compared to the case in Fig.~\ref{fig:shallow1}.
Color scheme is identical to Fig.~\ref{fig:shallow1}.
The initial
configuration in (a) is obtained by multiplying 0.5 to the random
initial $s_x(\vec{i})$ and $s_y(\vec{i})$ in
Fig.~\ref{fig:shallow1}(a).
The comparison with Fig.~\ref{fig:shallow1} indicates that
the low temperature phase depends on how the system is prepared,
as observed in some manganites.
}
\end{figure}

Figure~\ref{fig:shallow2} shows results of a simulation with
parameters identical to those in Fig.~\ref{fig:shallow1} except for a
narrower range of the random initial values of $s_x(\vec{i})$ and
$s_y(\vec{i})$ between $-s_0$ and $s_0$. Instead of multiple
nucleations, only one nucleation emerges within the $32\times32$
lattice, which grows and evolves the whole system into a periodic
patten of stripes with positive and negative $p_3$. The final state
shown in Fig.~\ref{fig:shallow2}(f) is another metastable phase not
considered in Fig.~\ref{fig:energy}, where distortions only with a wave
vector $(\pi,\pi)$ are considered. The result shows that
multiple inequivalent metastable phases exist even in this simple
model, and the coexistence of more than two phases is possible, as
suggested in some manganites.~\cite{Lee02}
We use an even narrower range of
the random initial $s_x(\vec{i})$ and $s_y(\vec{i})$ between $-s_0/2$
and $s_0/2$, in which case the system fails to nucleate a distorted
region and remains in the undistorted phase, showing the
characteristic metastability of systems with a first-order-transition energy
landscape.
The above results indicate that the low temperature metastable configurations
depend sensitively on how the configurations are obtained,
consistent with path-dependent experiments in manganites, such as
the sensitivity to the cooling rate or strain glass behaviors.~\cite{Wu06}

\begin{figure}[ht]
\leavevmode \epsfxsize15cm\epsfbox{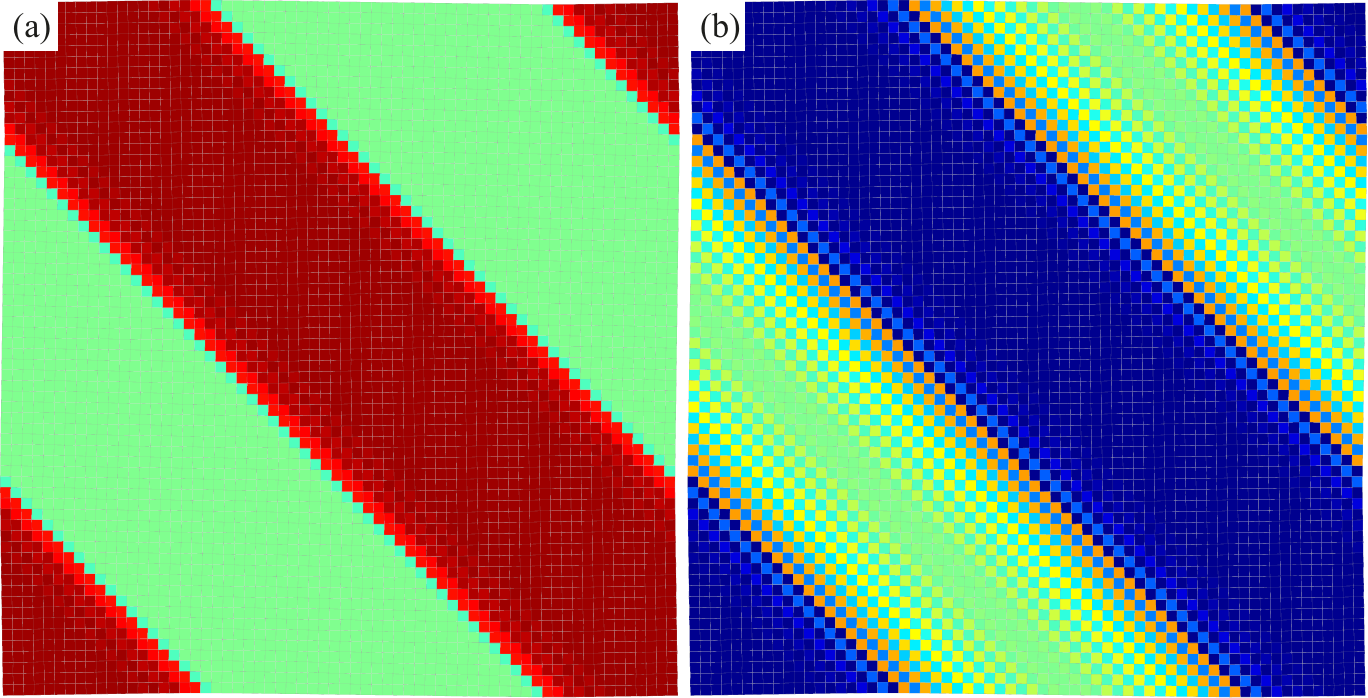}
\caption{\label{fig:deep_lat_ldos} (Color online)
(a) Stable configuration of
distorted and undistorted domains for a $64 \times 64$ lattice
for the energy landscape with a deep local minimum,
shown by the thick red curve in Fig.~\ref{fig:energy}(b). The color represents
$p_3$ with red and green for $s_o^2$ and 0, respectively.
(b) Map of
the local electron DOS calculated at $E=0$ for the distortions in (a). Red,
green, and blue correspond to 0.6, 0.3, and 0 state per site
per energy, respectively.
There is no intrinsic length scale for the size of the domains in our model,
and the size of the domain depends only on the predesigned initial condition and
can be as large as micron in principle,
which is consistent with experimental observations.
}
\end{figure}

For the deep local energy minimum case described by the thick red
curve in Fig.~\ref{fig:energy}(b), the simulation of rapid quenching
using the Euler method for a $64\times 64$ lattice does not create
nucleation of the low energy phase. Instead, we always obtain the undistorted
homogeneous state as the final state, which is an indication of
strong metastability due to a higher energy barrier between
the distorted and the undistorted states. In crystals, we expect line or
planar defects, as well as thermal fluctuations,
would assist nucleation. Simulations of such processes require more
computational resources. Therefore, we start from a predesigned
initial condition and relax the lattice to obtain stable coexistence
of distorted and undistorted domains. The initial condition is
chosen on a $64\times64$ lattice according to
\begin{eqnarray}
s_x(i_x,i_y)&=&s_0(-1)^{i_x+i_y}
\left\{\cos\left[\frac{2\pi(i_x+i_y-4)}{N}\right]+0.5\right\}, \\
s_y(i_x,i_y)&=&0,
\end{eqnarray}
where $N=64$. The initial
configuration is relaxed with $\gamma=0.0002$, and the stable
configuration is obtained, which is shown in
Fig.~\ref{fig:deep_lat_ldos}(a). We find stable coexistence of large
undistorted [green region in Fig.~\ref{fig:deep_lat_ldos}(a)] and
distorted [red region in Fig.~\ref{fig:deep_lat_ldos}(a)] domains,
unlike the shallow local minimum case studied above.  The size of
the domain is determined only by the initial condition, and
therefore can be as large as several micrometers, consistent with
experiments for manganites.

For comparison, we carry out simulations
for a $N \times N$ system with a short range interaction only, described by
$E_M$ in Eq.~(\ref{eq:EM}) with the parameter $W=2.0$, with
a similar predesigned initial condition,
\begin{equation}
M(i_x,i_y)=M_0 \left\{\cos\left[\frac{2\pi(i_x+i_y)}{N}\right]+c_{s}\right\}.
\end{equation}
For $c_s=0.5$, $\gamma=0.0002$, and $N=64$,
we find that the final stable configuration is
a uniform ground state for this system with a short interaction only,
rather than a state with domains.
For $c_s=0$, $\gamma=0.0002$, and $N=128$,
we find only a line of atoms, rather than a domain, with $M$
close to zero between regions with $M = M_0$ and $M = -M_0$.
This comparison shows that the strain-strain long range interaction indeed plays
an essential role for the coexistence of distorted and undistorted
phases.

\begin{figure}[ht]
\includegraphics[scale=0.5, clip=true]{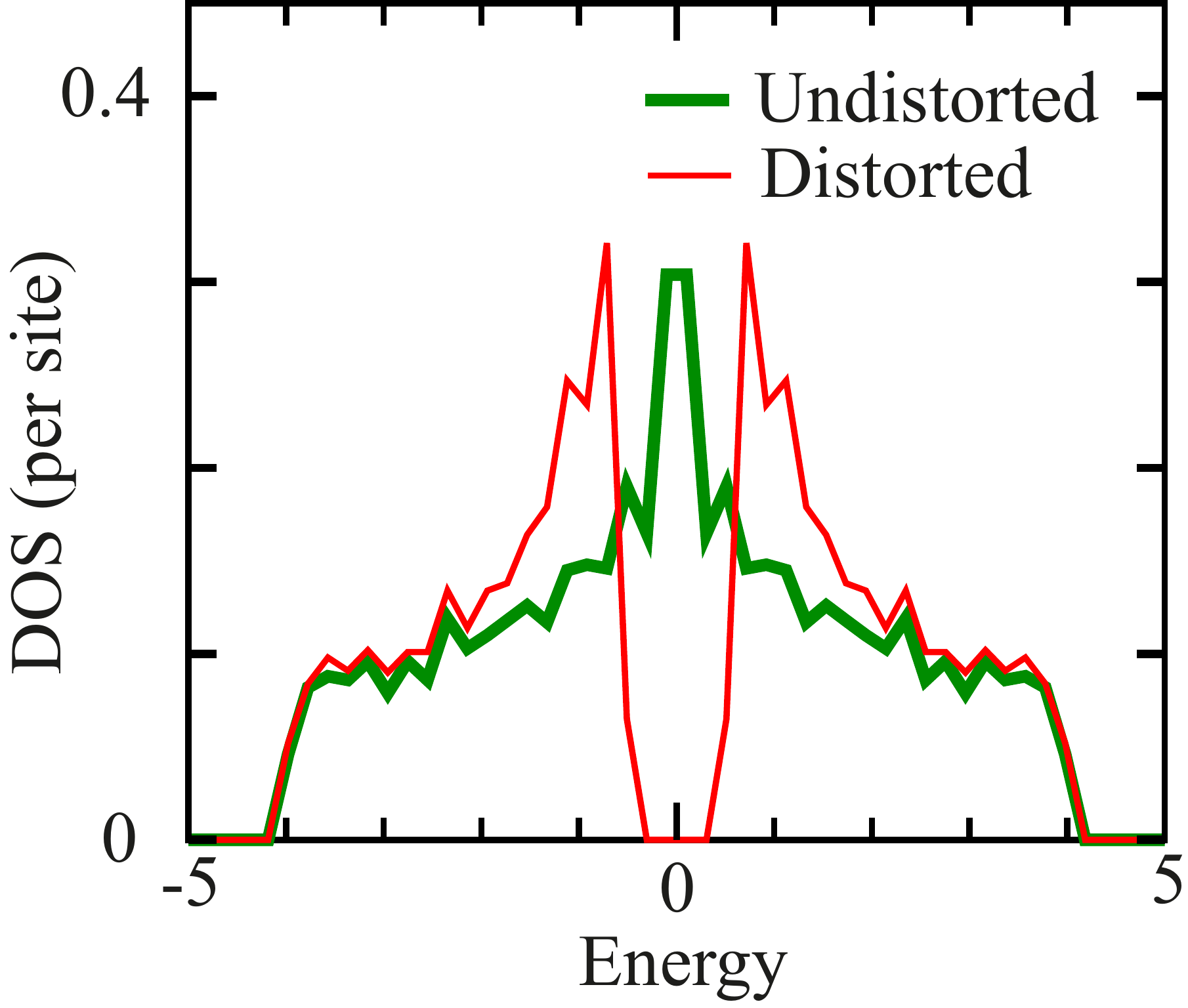}
\caption{\label{fig:deep_ldos_mi} (Color online)
Local electron DOS calculated at the center
of an undistorted region (thick green line), $\vec{i}=(36,1)$ in Fig.~\ref{fig:deep_lat_ldos}(a),
 and at the center of
a distorted region (thin red line), $\vec{i}=(6,1)$ in Fig.~\ref{fig:deep_lat_ldos}(a).
Unlike the case of small domains in Fig.~\ref{fig:ldos_shallow}, a clear gap opens
inside the large domain of distorted phase. The local density of states is always symmetric with respect to $E=0$.
Therefore, for $E_F=0$, the electron number is 0.5 at all sites,
demonstrating that the metal-insulator domain structure does not requires charge inhomogeneity.
}
\end{figure}

For the configuration in Fig.~\ref{fig:deep_lat_ldos}(a),
the local DOS versus energy is
calculated at the centers of the undistorted region
and the distorted region, which is shown in
Fig.~\ref{fig:deep_ldos_mi}. The local DOS within the undistorted
regions shows a metallic DOS without a gap, whereas the
local DOS within the distorted region shows an insulating DOS with a
gap around $E=0$. Therefore, for the chemical potential chosen at
$E=0$ inside this gap, we obtain the coexistence of metallic undistorted
and insulating distorted regions, as shown in
Fig.~\ref{fig:deep_lat_ldos}(b), similar to experiment results.
We also find that the interface between the metallic and
insulating regions is rather sharp, consistent with STM
images of atomically sharp interface between metallic and insulating domains in
manganites.~\cite{Renner02} Our results indicate that chemical
inhomogeneity is not a necessary condition to have a large scale
coexistence of metallic and insulating domains, which is
in contrast to other theories.~\cite{Burgy04} Although the
lattice defects or segregation of dopants could play a role in
nucleation, the stability of coexistence relies on the intrinsic
energy landscape, which explains why external perturbations such as
focused x-rays,~\cite{Kiryukhin97} light,~\cite{Fiebig98} or electron
beams alter metallic and insulating domains.

\section{Stability of phase coexistence}
\subsection{Stability against uniform domain wall motions}
\begin{figure}[ht]
\includegraphics[scale=1.0, clip=true]{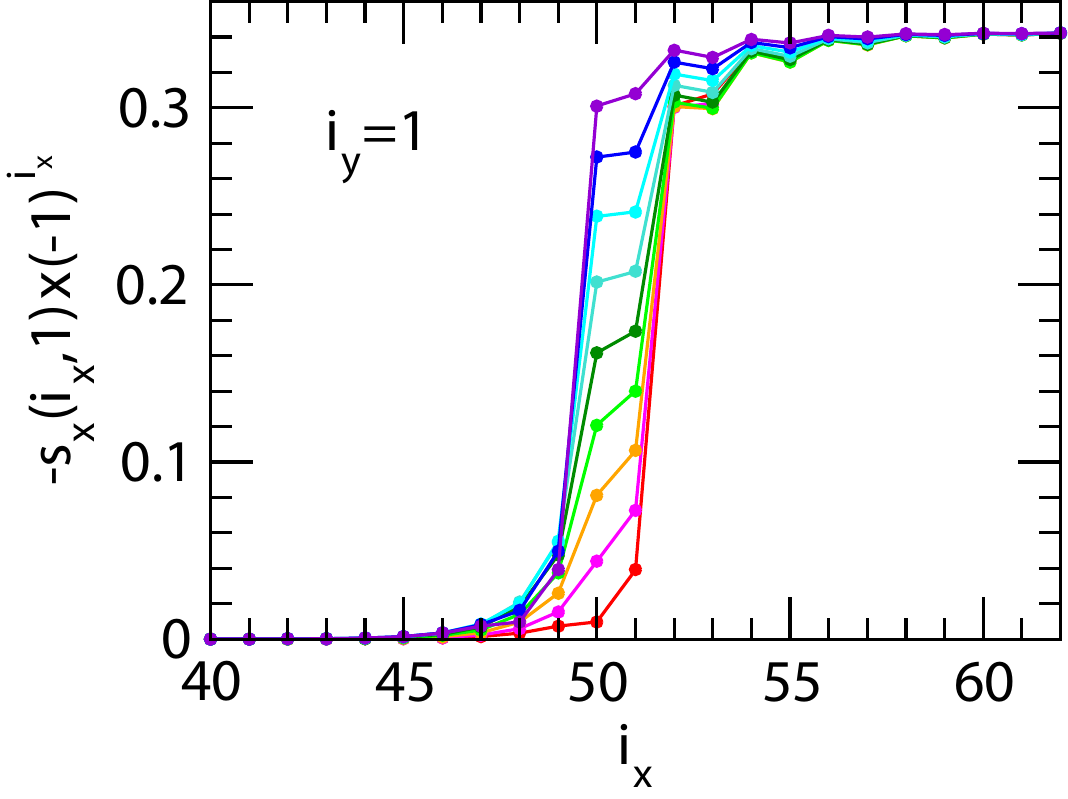}
\caption{\label{fig:unif_profile}  (Color online)
The red dots represent
the profiles of
$s_x(i_x,i_y)\times(-1)^{i_x+i_y}$ with $i_y=1$ near the domain
boundary in Fig.~\ref{fig:deep_lat_ldos}(a).
Other dots show how this profile changes as the domain boundary shifts uniformly
by two interatomic distances.
Lines are drawn to guide eyes.
The final configuration shown in purple dots is equivalent to a parallel shift
of the initial configuration shown in red dots, but the intermediate configurations are not,
due to the discreteness of the lattice.
}
\end{figure}

\begin{figure}[ht]
\includegraphics[scale=2.0, clip=true]{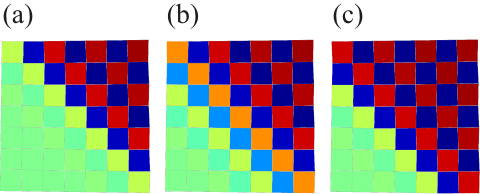}
\caption{\label{fig:unif_map} (Color online)
Configuration of lattice distortions
near the domain boundary, as the boundary moves by two interatomic
distances. The color scheme is different from Figs.~\ref{fig:shallow1}, \ref{fig:shallow2}, and \ref{fig:deep_lat_ldos}(a).
Here, the color represents the $s_x$ mode, with red, green, and blue
corresponding to $s_o$, 0, and $-s_o$, respectively.
(a), (b), and (c) correspond to the profile represented by red, green, and purple dots in
Fig.~\ref{fig:unif_profile}.
Configuration (c) has a lower energy than (a),
but the higher energy of the intermediate configuration (b) serves as a Peierls-Nabarro barrier,
as shown in Fig.~\ref{fig:unif_energy}.
}
\end{figure}

In this section, we examine the stability of the phase coexistence against various kinds of
perturbations. First, we examine the energy barrier blocking a uniform
shift of the domain boundaries, which would convert the undistorted
high energy phase into the distorted low energy phase. Red dots
connected by the lowest lines in Fig.~\ref{fig:unif_profile} show
$s_x(i_x,i_y)\times(-1)^{i_x+i_y}$ versus $i_x$ for $i_y=1$
near the boundary between the undistorted (i.e., $i_x\leq 51$) and
distorted (i.e., $i_x \geq 52$) phases for the configuration in
Fig.~\ref{fig:deep_lat_ldos}(a). To find the energy barrier against
uniform domain wall shift, we increase the value of
$s_x(i_x,i_y)\times(-1)^{i_x+i_y}$ at the sites immediately adjacent to the
domain boundary, that is, at $i_x=51-i_s$, $i_y=1+i_s$ with integer
$i_s$'s, in 8 steps from near zero to the full distortion close to
$s_0$. At each step, we minimize the total energy with respect to the distortions
at all other sites using the Euler method. This gives rise to the
distortion profiles along the horizontal direction shown in
Fig.~\ref{fig:unif_profile}. The 2D configurations for
the red, green, and purple dots in Fig.~\ref{fig:unif_profile} are also shown
in Figs.~\ref{fig:unif_map}(a), \ref{fig:unif_map}(b), and \ref{fig:unif_map}(c), respectively,
where the color represents $s_x$. The results show that $s_x(i_x,i_y)\times(-1)^{i_x+i_y}$ at
$(51,1)$ and $(50,1)$ grow together, compensating
$s_x$  distortions with opposite signs at the two neighboring sites,
and the domain boundary advances by two interatomic distances.

\begin{figure}[ht]
\includegraphics[scale=1.0, clip=true]{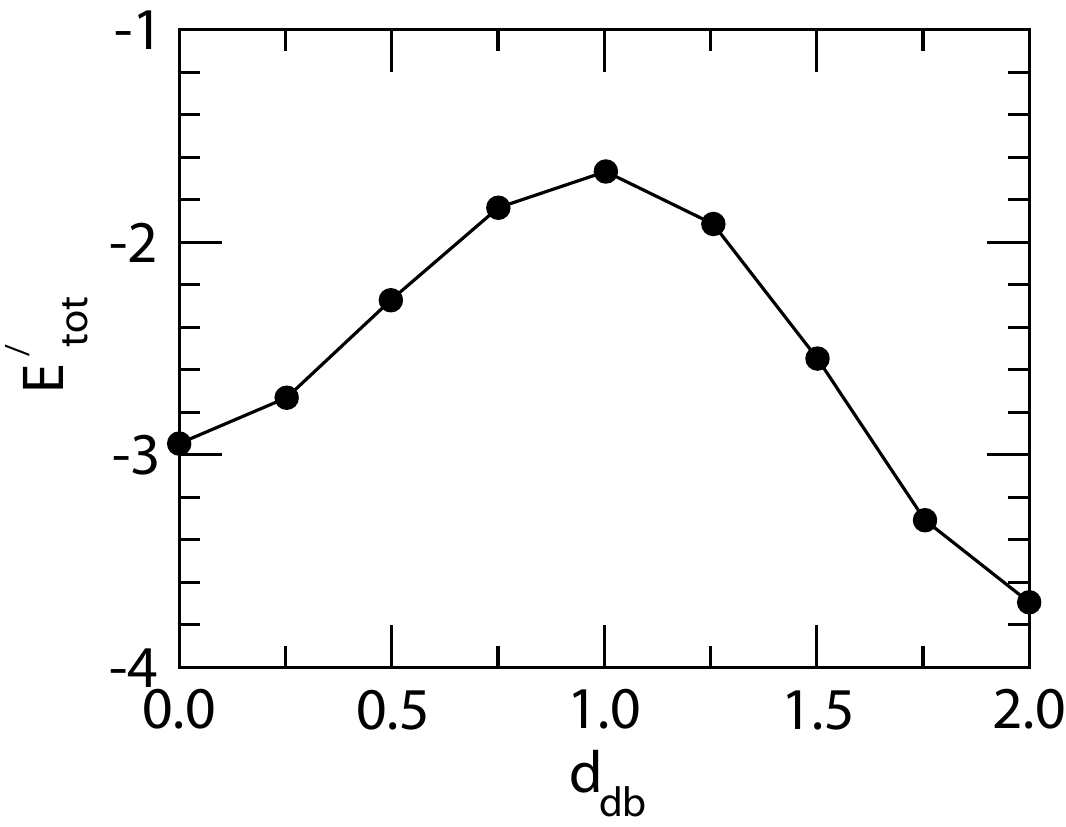}
\caption{\label{fig:unif_energy} Total energy $E'_{tot}$ for the 64 $\times$ 64 system, given by
Eq.~(\ref{eq:Eptot}), versus the location of the domain boundary
defined by Eq.~(\ref{eq:idb}), as the boundary moves by two
interatomic distances. Each point is found from each corresponding curve in
Fig.~\ref{fig:unif_profile}. The configurations in
Figs.~\ref{fig:unif_map}(a), \ref{fig:unif_map}(b), and
\ref{fig:unif_map}(c) correspond to $d_{db}=0.0$, 1.0, and 2.0,
respectively.
The energy barrier prevents a parallel shift of the domain wall.
}
\end{figure}

We define the effective location of the domain boundary, $d_{db}$,
according to
\begin{equation}
d_{db}=2\times\frac{s_x^*-0.040}{0.308-0.040}, \label{eq:idb}
\end{equation}
where $s_x^*$ represents the value of
$s_x(i_x,i_y)\times(-1)^{i_x+i_y}$ chosen at $i_x=51$ in Fig.~\ref{fig:unif_profile},
and 0.040 and 0.308  the values
of $s_x^*$ before and after the domain wall moves by two interatomic
distances. The minimized total energy $E'_{tot}$
is plotted in Fig.~\ref{fig:unif_energy} for $0 \leq d_{db} \leq 2$,
which shows the energy barrier.
We compare three energies, $E_1=-2.946$, $E_2=-1.668$, and
$E_3=-3.693$ for $d_{db}$ = 0.0, 1.0, and 2.0, which correspond to the
configurations shown in Figs.~\ref{fig:unif_map}(a),
\ref{fig:unif_map}(b), and \ref{fig:unif_map}(c), respectively. Most
changes in distortion occur in the $64\times2$ plaquettes near the
domain boundary, as shown in Figs.~\ref{fig:unif_profile} and
~\ref{fig:unif_map}. Therefore, the energy difference between the
two stable domain configurations for $d_{db}=$ 0.0 and 2.0 is $(E_1-E_3)/128=0.0058$
per site, which agrees with the energy difference per site, 0.0058, between
the undistorted and distorted uniform phases in
Fig.~\ref{fig:energy}(b). The energy barrier normalized for
$64\times 2$ plaquettes, that is, $(E_2-E_1)/128$, is 0.0100, which
is of the same order of magnitude as the height of the energy
barrier, $\Delta E=0.0160$, between the local and global energy
minima in Fig.~\ref{fig:energy}(b). From this analysis, the energy
barrier against the uniform shift of the domain wall would be of the
order of $2 \Delta E$ multiplied by the domain wall length in the
units of interatomic distance, which would be a macroscopic energy
barrier for the domain walls of micron length scale.
We emphasize that discreteness of the lattice in
our model is essential for this energy barrier, which is an example
of the Peierls-Nabarro barrier.~\cite{Nabarro}

\subsection{Stability against non-uniform domain wall modifications}
\begin{figure}[ht]
\leavevmode \epsfxsize15cm\epsfbox{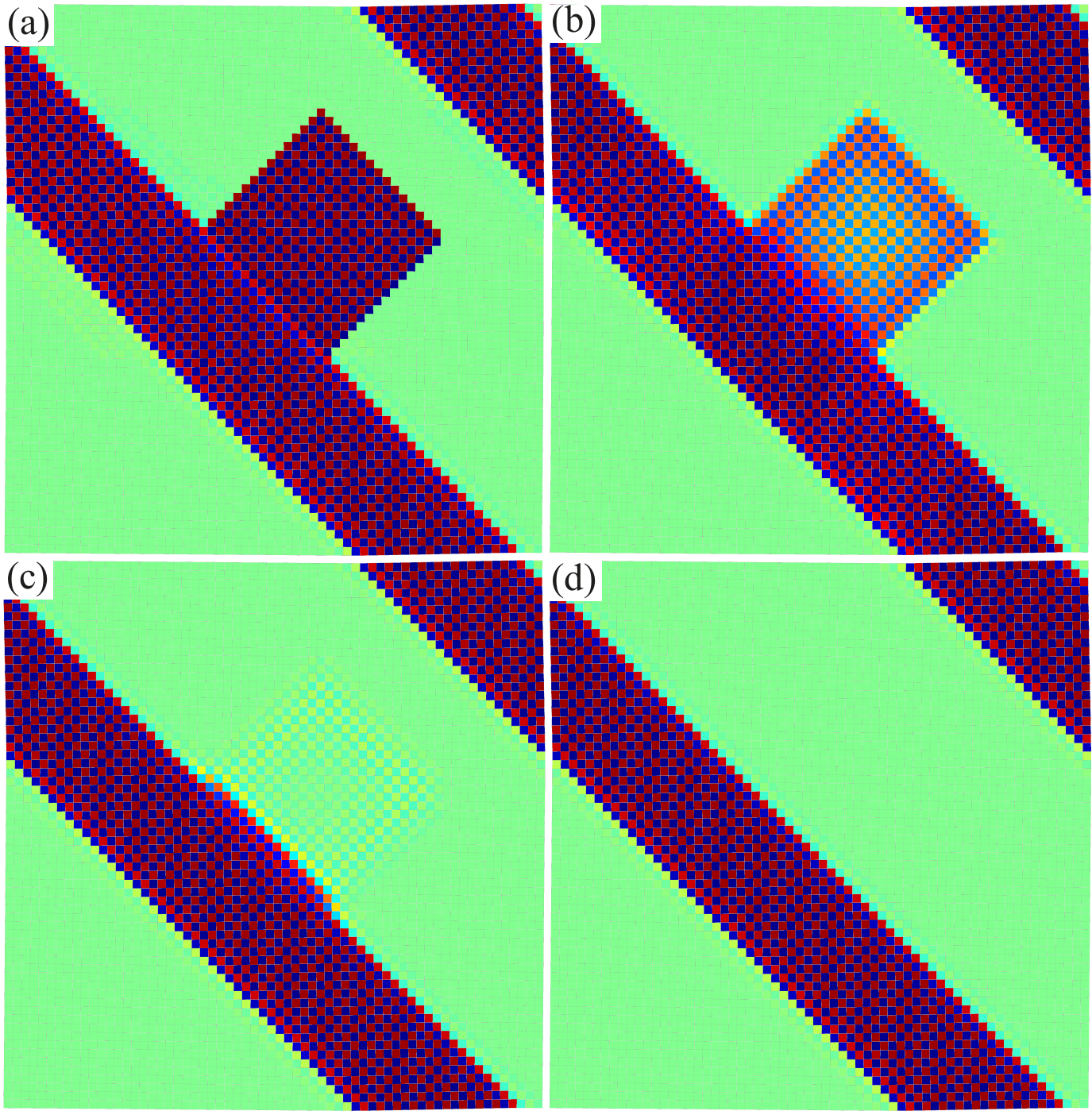}
\caption{\label{fig:nonunif_sd} (Color online)
Simulation of the domain wall
stability against a small nonuniform modification of the domain boundary
for the configuration similar to Fig.~\ref{fig:deep_lat_ldos}(a).
The color scheme is identical to
Fig.~\ref{fig:unif_map}.
(a) represents the initial perturbed configuration. (b)
and (c) show intermediate configurations. (d) represents the final
stable configuration, which is identical to the original
configuration before the perturbation.
It shows that the original domain configuration,
such as the one shown in Fig.~\ref{fig:deep_lat_ldos},
is stable against
small non-parallel shift of the domain boundary.
}
\end{figure}

\begin{figure}[ht]
\leavevmode \epsfxsize15cm\epsfbox{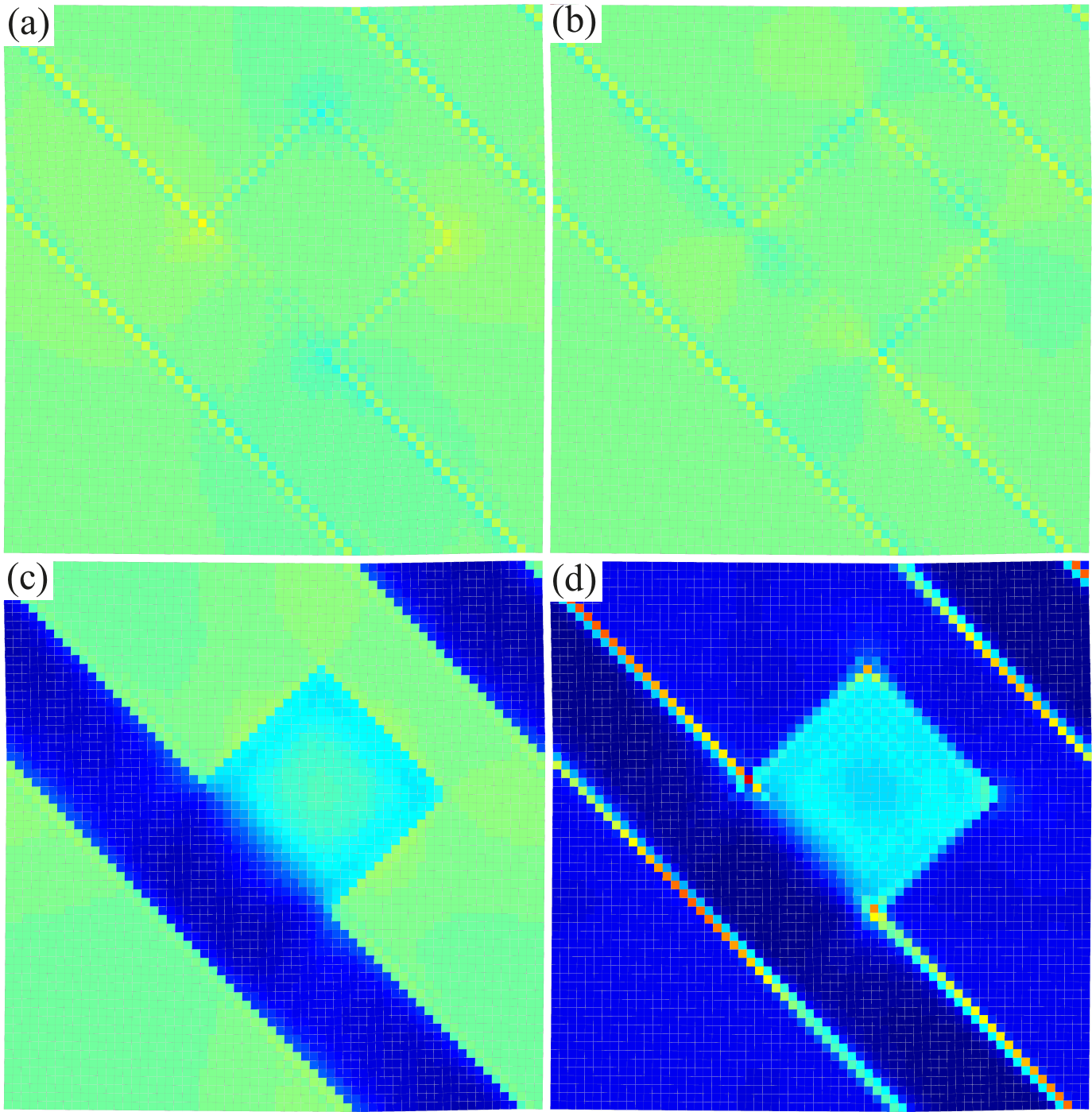}
\caption{\label{fig:nonunif_sd_modes} (Color online)
Colors in (a), (b) and (c) show the
$e_1(\vec{i})$, $e_2(\vec{i})$, and $e_3(\vec{i})$ for the configuration shown in
Fig.~\ref{fig:nonunif_sd}(b).
Colors in (d) show $E_{tot}(\vec{i})$, which is
the sum of the terms with the site index $\vec{i}$
in Eqs.~(\ref{eq:Es})-(\ref{eq:Ec}).
In the panels (a), (b), and (c), red and
blue correspond to $\pm 0.45$ and green to zero. In the panel
(d), red and blue correspond to 0.06 and -0.006, respectively.
Typical values of $e_3(\vec{i})$ and $E_{tot}(\vec{i})$ inside the converted patch
are -0.08 and 0.02, respectively.
Panel (d) shows that the energy cost for changing domain wall in a non-parallel way
spreads over the whole changed area (volume in 3D), which is different
from systems with a short range interaction only and is responsible for the stability
against such non-parallel shift of the domain walls.
}
\end{figure}

The importance of the long range interaction between strain fields
is even more evident for the stability against nonuniform modification of
domain walls. As an example, we convert a patch of the undistorted
region in a configuration similar to Fig.~\ref{fig:deep_lat_ldos}(a),
into a distorted state initially and then relax the whole lattice according
to the Euler method. The initial, two intermediate, and final
configurations are displayed in Fig.~\ref{fig:nonunif_sd}. The results
show that the distortion in the converted region disappears
initially except for two atomic layers [Fig.~\ref{fig:nonunif_sd}(c)],
which shrink laterally by further relaxation, restoring the original
configuration [Fig.~\ref{fig:nonunif_sd}(d)]. The simulation
demonstrates the stability of domain structure against non-uniform
modification of the domain walls. To gain further insight into the
role of lattice compatibility, we examine other modes and energy
distributions. Figures~\ref{fig:nonunif_sd_modes}(a),
\ref{fig:nonunif_sd_modes}(b) and \ref{fig:nonunif_sd_modes}(c) show
the modes, $e_1$, $e_2$, and $e_3$ for the $s_x$ distortion given in
Fig.~\ref{fig:nonunif_sd}(b). The strain field tends to spread into
the domains from the domain boundary. In particular, the $e_3$ field
inside the converted patch in Fig.~\ref{fig:nonunif_sd_modes}(c)
cannot reach $-C_3s_0^2/A_3$, the full distortion of $e_3$ inside
the domain, due to the strain compatibility.
The map of $E_{tot}(\vec{i})$, the sum of the terms with the site index $\vec{i}$
in Eqs.~(\ref{eq:Es})-(\ref{eq:Ec}), is
shown in Fig.~\ref{fig:nonunif_sd_modes}(d), which implies that the energy cost
for creating the distorted patch is not confined immediately around the
interface, but is distributed over the whole converted patch. This is
different from systems with short range interactions only, for which the
energy cost would be confined near the domain wall within the range of the interaction.
This difference shows that the lattice
constraint, leading to the effective long range interaction, plays an
important role in the stability of phase coexistence against non-uniform modification
of the domain boundary. Similarly, we
find that if we convert a patch of the distorted region of a
similar size as above into the undistorted phase, the system
relaxes back to the original configuration.

\begin{figure}[ht]
\leavevmode \epsfxsize15cm\epsfbox{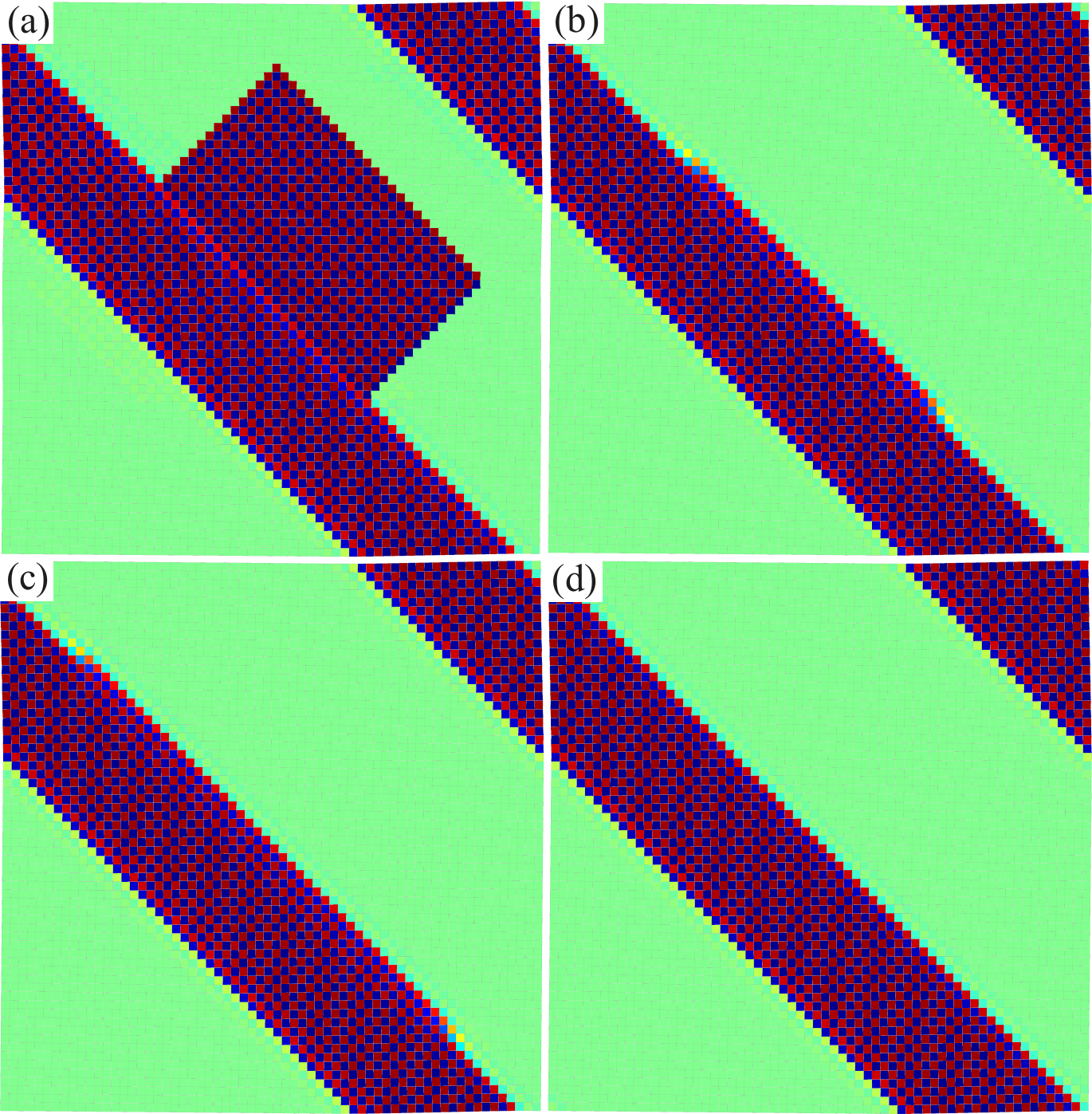}
\caption{\label{fig:nonunif_ld} (Color online)
Simulation of the domain wall shift by
large enough nonuniform modification of the domain boundary,
converting a large undistorted patch into a distorted state, for
the configuration similar to Fig.~\ref{fig:deep_lat_ldos}(a).
The color scheme is identical to Fig.~\ref{fig:unif_map}.
An initial configuration is shown in (a).
(b) and (c) show intermediate configurations. (d)
represents the final stable configuration, which shows that the
distorted region has expanded by two atomic layers.
It shows that if the perturbation is large enough, the system relaxes
to a different metastable configuration, unlike the case in Fig.~\ref{fig:nonunif_sd}.
}
\end{figure}

\begin{figure}[ht]
\leavevmode \epsfxsize15cm\epsfbox{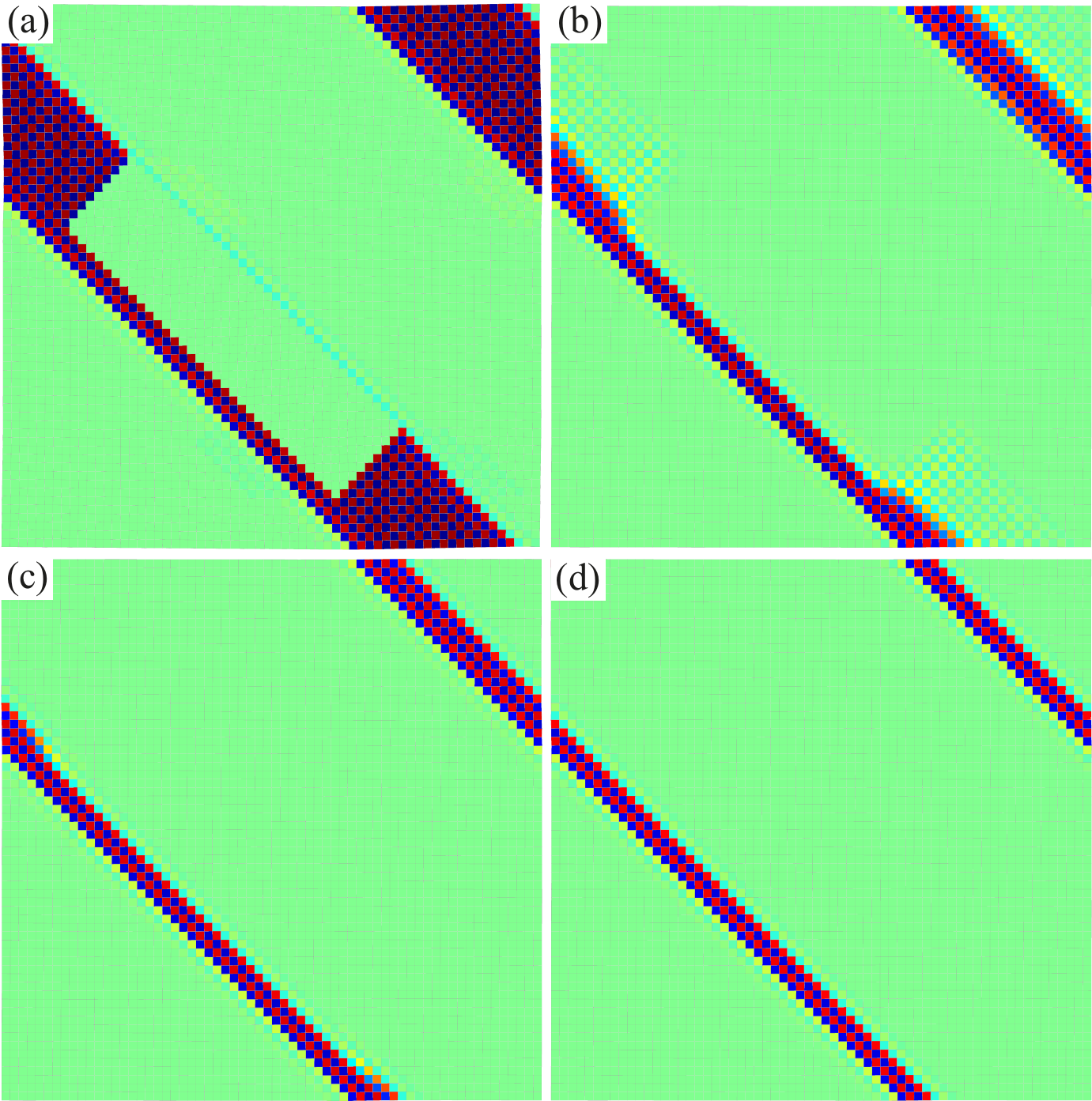}
\caption{\label{fig:nonunif_lud} (Color online)
Simulation of the domain wall shift
by a large enough nonuniform modification of the domain boundary,
converting a large distorted patch into an undistorted state,
for the configuration similar to Fig.~\ref{fig:deep_lat_ldos}(a).
The color scheme is identical to Fig.~\ref{fig:unif_map}.
An initial configuration is shown in (a).
(b) and (c) show intermediate configurations. (d) represents
the final stable configuration, which shows that
the undistorted metallic region has expanded.
This simulation may be related to the experiment results in Ref.~\onlinecite{Kiryukhin97},
in which the x-rays destroy the charge-orbital ordering and
the short wavelength distortions and increase conductivity.
}
\end{figure}

However, the above results do not mean that it is impossible to
convert a region between phases. For example, if
we convert a large enough patch, as shown in
Fig.~\ref{fig:nonunif_ld}(a), even though the distortion in the most
converted region disappears initially, the two distorted layers remaining
 in Fig.~\ref{fig:nonunif_ld}(b) expand laterally, as shown in
Fig.~\ref{fig:nonunif_ld}(c). Eventually, the distorted domain
grows by two atomic layers, as shown in Fig.~\ref{fig:nonunif_ld}(d).
These results, particularly the different relaxation behavior for the configurations
in Figs.~\ref{fig:nonunif_sd}(c) and \ref{fig:nonunif_ld}(c),
show that the energy barrier for the growth of the low energy phase
involves simultaneous distortions of a significant number of unit cells just next to
the domain wall.
Slow growth of the low energy phase has been observed in a number of experiments
for manganites. For example, Ref.~\onlinecite{Tao05} reports a time scale of the order of 10 minutes
for the growth of the low energy phase.
A rough order of magnitude estimate of the energy barrier $\Delta E$ can be made
by assuming an activated thermal process so that  the relaxation time
$\tau = \tau_0 \exp (\Delta E/k_B T)$,
where $\tau_0$ represents the intrinsic time scale for ion motion,
$k_B$ is the Boltzmann constant, and $T$ is temperature.
With $\tau \sim 10^3$ s, $\tau_0 \sim 10^{-13}$ s, $k_B T \sim 10$ meV,
we obtain $\Delta E$ of the order of 1 eV.
If we consider the typical energy scale for the distortion of unit cell to be 1 - 10 meV,
this energy barrier corresponds to about 100 - 1000 unit cell distortions
within the layer just next to the 2-dimensional interface,
consistent with our simulations.
A similar growth of the undistorted region
occurs if we convert a large enough distorted region into the
undistorted phase, as shown in Fig.~\ref{fig:nonunif_lud}.
The result in Fig.~\ref{fig:nonunif_lud}
is reminiscent of experiments in which the volume fraction
of the undistorted phase is increased by external perturbations such as
x-rays or light.~\cite{Kiryukhin97,Fiebig98}

\section{Summary}
We have discussed various aspects of a model for the strain-induced phase
coexistence observed in perovskite manganites. A square lattice
and associated atomic scale distortion modes were used to construct
an energy expression with local and global energy minimum states,
which captures features of
manganites essential for  phase coexistence:
a local energy minimum metallic state without lattice distortions
and a global energy minimum insulating state with short wavelength and uniform lattice
distortions. Explicit expressions for modes, constraint equations,
energies, and energy gradients have been presented.
Our simulations for an energy landscape with a low energy barrier against
transforming from undistorted local to distorted global energy minimum states
revealed nucleation with anisotropic correlation
upon rapid quenching.
Our simulations for an energy landscape with a high energy barrier
showed stable coexistence of undistorted metallic and distorted insulating domains.
Further, we studied the stability of such metal-insulator
domain structures against various perturbations. We found that domain
configurations are stable against uniform motion of the boundary due
to the discreteness of the lattice and the intrinsic energy barrier
between local and global energy minimum states. We expect  that this intrinsic atomic
scale energy barrier, multiplied by the number of atoms within the
mesoscopic scale domain wall, is  large enough to
prevent the uniform motion of domain walls. For non-uniform modification
of these walls, the long range interaction between strain fields
gives rise to the domain wall energy distributed over the whole
modified area for our 2D model (or volume for 3D
system), rather than just the region confined near the domain
wall, providing extra stability to the domain structure.
To provide comparison, we carried out simulations
for a system with a short range interaction only,
which show no anisotropic nucleation or stable coexistence of
local and global energy minimum phases.
The above results demonstrate that
the long range interaction between stain fields, and associated complex energy landscape,
play an important role in metal-insulator coexistence in perovskite manganites.

Establishment of more concrete connections between our model and experiments
would be the goal of future studies.
For example, the density of states at the Fermi energy level
in the inhomogeneous state can be compared with the conductivity measured in experiments.
The effect of substrate-induced strain can be simulated
in our model with additional energy terms
representing the bonding between atoms in the film and the substrate.
Thermal fluctuation can be simulated by the Monte Carlo method,
which may provide insights into the origin of ``strain" glass behavior
that has been experimentally proposed as intrinsic rather than extrinsic.~\cite{Wu06}
Furthermore, although our framework is based on the assumption that
electronic and magnetic effects are adiabatically slaved to lattice distortions,
our work can, in principle, be generalized to include these functionalities
in a self-consistent manner. Such coupled models
will be computationally intensive and our approach has been to seek a minimal model.
However, discrete strain or pseudo-spin models~\cite{Lookman08} with long-range interactions
and disorder provide a skeletal approach to couple with magnetic spins
and electronic densities within a mean-field or Monte Carlo scheme. Here the abundant
literature on spin models is an advantage because even glassy behavior
in electronic materials may be identified by an appropriate order-parameter.

\section{Acknowledgement}
We thank Avadh Saxena for discussions.
This work has been supported by U.S. Department of Energy and NJIT.

\appendix

\section{Energy expressions for inhomogeneous states}

First, we represent $e_1$, $e_2$, and $e_3$ in the reciprocal space,
and rewrite $E_{l}$ and $E_{c}$  in Eqs.~(\ref{eq:El}) and
(\ref{eq:Ec}) in the following form:
\begin{eqnarray}
E_{l}&=&N^2\sum_{\vec{k}}
\frac{A_1}{2}  e_1(\vec{k}) e_1(-\vec{k})
+\frac{A_2}{2}  e_2(\vec{k}) e_2(-\vec{k})
+\frac{A_3}{2}  e_3(\vec{k}) e_3(-\vec{k}), \label{eq:Elk} \\
E_{c}&=&\sum_{\vec{i}} \left\{ C_3 \left[s_x(\vec{i})^2-s_y(\vec{i})^2\right]
\sum_{\vec{k}} e_3(\vec{k}) e^{i\vec{k}\cdot\vec{i}} \right\}. \label{eq:Eck}
\end{eqnarray}
Next, because the constraint equations apply differently depending
on whether either $k_x$ or $k_y$ is zero or not, we divide the
$\vec{k}$-sum into four parts,
\begin{equation}
\sum_{\vec{k}}=\sum_{k_x\neq0,k_y\neq0}+\sum_{k_x=0,k_y\neq0}+\sum_{k_x\neq0,k_y=0}+\sum_{k_x=0,k_y=0},
\end{equation}
and treat each of them separately. If $k_x \neq 0$ and $k_y \neq 0$,
constraint equations
Eqs.~(\ref{eq:constraint1})-(\ref{eq:constraint3}) are rewritten in
the following way, which expresses the modes $e_1(\vec{k})$,
$e_2(\vec{k})$, $e_3(\vec{k})$ in terms of $s_x(\vec{k})$ and
$s_y(\vec{k})$:
\begin{eqnarray}
(e_1)_{k_x\neq0,k_y\neq0}&=&-\frac{i}{\sqrt{2}} \left[ \cot \frac{k_y}{2} s_x(\vec{k}) + \cot \frac{k_x}{2} s_y(\vec{k}) \right],
\label{eq:e1kxn0yn0} \\
(e_2)_{k_x\neq0,k_y\neq0}&=&-\frac{i}{\sqrt{2}} \left[ \cot \frac{k_x}{2} s_x(\vec{k}) + \cot \frac{k_y}{2} s_y(\vec{k}) \right],
\label{eq:e2kxn0yn0} \\
(e_3)_{k_x\neq0,k_y\neq0}&=&-\frac{i}{\sqrt{2}} \left[ \cot \frac{k_y}{2} s_x(\vec{k}) - \cot \frac{k_x}{2} s_y(\vec{k}) \right].
\label{eq:e3kxn0yn0}
\end{eqnarray}
Therefore, the part with $k_x \neq 0$ and $k_y \neq 0$ for $E_{l}$
in Eq.~(\ref{eq:Elk}) is expressed as
\begin{equation}
(E_{l})_{k_x\neq0,k_y\neq0}
= \frac{N^2}{2}
\sum_{k_x\neq0,k_y\neq0}
\left(
\begin{array}{c}
s_x \\
s_y
\end{array}
\right)^T_{-\vec{k}}
\left(
\begin{array}{cc}
B_{xx}  & B_{xy}  \\
B_{xy}  & B_{yy}
\end{array}
\right)_{\vec{k}}
\left(
\begin{array}{c}
s_x \\
s_y
\end{array}
\right)_{\vec{k}},
\end{equation}
where
\begin{eqnarray}
B_{xx}(\vec{k})&=&\frac{1}{2}(A_1+A_3) \cot^2 \frac{k_y}{2} + \frac{1}{2} A_2 \cot^2 \frac{k_x}{2}, \\
B_{yy}(\vec{k})&=&\frac{1}{2}(A_1+A_3) \cot^2 \frac{k_x}{2} + \frac{1}{2} A_2 \cot^2 \frac{k_y}{2}, \\
B_{xy}(\vec{k})&=&\frac{1}{2}(A_1+A_2-A_3) \cot \frac{k_x}{2} \cot \frac{k_y}{2}.
\end{eqnarray}
Similarly, the part with $k_x\neq0$ and $k_y\neq0$ for $E_{c}$ in
Eq.~(\ref{eq:Eck}) is equivalent to
\begin{equation}
(E_{c})_{k_x\neq0,k_y\neq0}=\sum_{\vec{i}} C_3 \left[ s_x(\vec{i})^2-s_y(\vec{i})^2 \right]
\sum_{k_x\neq0,k_y\neq0} -\frac{i}{\sqrt{2}} \left[ \cot \frac{k_y}{2} s_x(\vec{k})
- \cot \frac{k_x}{2} s_y(\vec{k}) \right] e^{i\vec{k}\cdot\vec{i}}.
\end{equation}
For the terms with $k_x = 0$ and $k_y \neq 0$, we apply the constraint
equation $e_1(\vec{k})+e_3(\vec{k})=0$ to eliminate $e_1(\vec{k})$
in $E_l$ in Eq.~(\ref{eq:Elk}) and  obtain
\begin{eqnarray}
(E_l+E_c)_{k_x=0,k_y\neq0}&=&N^2 \sum_{k_x=0,k_y\neq0}
\frac{1}{2}(A_1+A_3) e_3(-\vec{k}) e_3(\vec{k}) +
\frac{1}{2} A_2 e_2(-\vec{k}) e_2(\vec{k}) \nonumber \\
& &+\sum_{\vec{i}} C_3 \left[s_x(\vec{i})^2-s_y(\vec{i})^2\right]
\sum_{k_x=0,k_y\neq0} e_3(\vec{k}) e^{i k_y i_y}.
\end{eqnarray}
Since we are interested in metastable phases in this work and
$e_2(\vec{k})$ and $e_3(\vec{k})$ are independent for $k_x = 0$ and $k_y \neq 0$,
we minimize the energy $(E_l+E_c)_{k_x=0,k_y\neq0}$ with respect to $e_2(\vec{k})$ and $e_3(\vec{k})$
separately and obtain
\begin{eqnarray}
&&(e_1)^{min}_{k_x=0,k_y\neq0}=-(e_3)^{min}_{k_x=0,k_y\neq0} = \frac{C_3}{A_1+A_3}
\mathcal{F}_{0,k_y}(s_x^2-s_y^2), \label{eq:e1kxe0yn0} \\
&&(e_2)^{min}_{k_x=0,k_y\neq0}=0, \label{eq:e2kxe0yn0}
\end{eqnarray}
where
\begin{equation}
\mathcal{F}_{\vec{k}}(s_x^2-s_y^2) \equiv \frac{1}{N^2}
\sum_{\vec{i}} \left[ s_x(\vec{i})^2- s_y(\vec{i})^2\right] e^{-i\vec{k}\cdot\vec{i}}.
\end{equation}
The minimized energy expression for $(E_l+E_c)_{k_x=0,k_y\neq0}$ is
\begin{equation}
(E_l+E_c)^{min}_{k_x=0,k_y\neq0} = - \frac{C_3^2}{2(A_1+A_3)} N^2
 \sum_{k_x=0,k_y\neq0} \mathcal{F}_{0,-k_y}(s_x^2-s_y^2)
\mathcal{F}_{0,k_y}(s_x^2-s_y^2).
\end{equation}
We apply a similar analysis for the terms with $k_x \neq 0$ and $k_y
= 0$ in Eqs.~(\ref{eq:Elk}) and (\ref{eq:Eck}). Using the constraint
$e_1(\vec{k})-e_3(\vec{k})=0$, we eliminate $e_1(\vec{k})$ and
obtain
\begin{eqnarray}
(E_l+E_c)_{k_x\neq0,k_y=0}&=&N^2 \sum_{k_x\neq0,k_y=0}
\frac{1}{2}(A_1+A_3) e_3(-\vec{k}) e_3(\vec{k}) +
\frac{1}{2} A_2 e_2(-\vec{k}) e_2(\vec{k}) \nonumber \\
& &+\sum_{\vec{i}} C_3 \left[s_x(\vec{i})^2-s_y(\vec{i})^2\right]
\sum_{k_x\neq0,k_y=0} e_3(\vec{k}) e^{i k_x i_x}.
\end{eqnarray}
Separate minimization of this energy with respect to $e_2(\vec{k})$
and $e_3(\vec{k})$ leads to
\begin{eqnarray}
(e_1)^{min}_{k_x\neq0,k_y=0}&=&(e_3)^{min}_{k_x\neq0,k_y=0} = -\frac{C_3}{A_1+A_3} \mathcal{F}_{k_x,0}(s_x^2-s_y^2), \label{eq:e1kxn0ye0} \\
(e_2)^{min}_{k_x\neq0,k_y=0}&=&0,  \label{eq:e2kxn0ye0} \\
(E_l+E_c)^{min}_{k_x\neq0,k_y=0}&=&- \frac{C_3^2}{2(A_1+A_3)} N^2
\sum_{k_x\neq0,k_y=0} \mathcal{F}_{-k_x,0}(s_x^2-s_y^2)
\mathcal{F}_{k_x,0}(s_x^2-s_y^2).
\end{eqnarray}
The terms with $\vec{k}=0$ in Eqs.~(\ref{eq:Elk}) and~(\ref{eq:Eck}) are
\begin{eqnarray}
(E_l+E_c)_{k_x=0,k_y=0}&=&
N^2 \left[ \frac{A_1}{2}  e_1(\vec{k}=0)^2  + \frac{A_2}{2}  e_2(\vec{k}=0)^2
+\frac{A_3}{2}  e_3(\vec{k}=0)^2 \right] \nonumber \\
& &+ \sum_{\vec{i}} C_3 \left[s_x(\vec{i})^2-s_y(\vec{i})^2\right] e_3(\vec{k}=0).
\end{eqnarray}
We minimize the above expression with respect to $e_1(\vec{k}=0)$,
$e_2(\vec{k}=0)$, and $e_3(\vec{k}=0)$ independently, since they are
not constrained to each other, and obtain
\begin{eqnarray}
(e_1)^{min}_{k_x=0,k_y=0}&=&0, \label{eq:e1kxe0ye0} \\
(e_2)^{min}_{k_x=0,k_y=0}&=&0, \label{eq:e2kxe0ye0} \\
(e_3)^{min}_{k_x=0,k_y=0}&=&
-\frac{C_3}{A_3}  \mathcal{F}_{\vec{k}=0}(s_x^2-s_y^2), \label{eq:e3kxe0ye0} \\
(E_l+E_c)^{min}_{k_x=0,k_y=0}
&=&-\frac{C_3^2}{2 A_3} N^2 \left[\mathcal{F}_{\vec{k}=0}(s_x^2-s_y^2)\right]^2.
\end{eqnarray}
Finally, by adding the terms with different cases of $k_x$ and $k_y$
found above, we obtain the following total energy, $E'_{tot}$, which
depends only on $s_x$ and $s_y$:
\begin{equation}
E'_{tot}(s_x,s_y)=E'_{l+c}+E_s, \label{eq:Eptot}
\end{equation}
where
\begin{eqnarray}
E'_{l+c}&=&(E_l)_{k_x\neq0,k_y\neq0}+(E_c)_{k_x\neq0,k_y\neq0}
 +(E_l+E_c)^{min}_{k_x\neq0,k_y=0}+(E_l+E_c)^{min}_{k_x=0,k_y\neq0} \nonumber \\
& &+(E_l+E_c)^{min}_{k_x=0,k_y=0},
\end{eqnarray}
and $E_s$ is given by Eq.~(\ref{eq:Es}). We use this energy
expression $E'_{tot}(s_x,s_y)$ for the simulations of inhomogeneous
states.

In addition to
$s_x(\vec{i})$ and $s_y(\vec{i})$ configurations, $e_1(\vec{i})$,
$e_2(\vec{i})$, and $e_3(\vec{i})$ configurations give useful information on the
nature of the inhomogeneous states. The relations used to eliminate
$e_1$, $e_2$, and $e_3$ variables above, namely,
Eqs.~(\ref{eq:e1kxn0yn0})-(\ref{eq:e3kxn0yn0}), (\ref{eq:e1kxe0yn0}), (\ref{eq:e2kxe0yn0}),
 (\ref{eq:e1kxn0ye0}), (\ref{eq:e2kxn0ye0}) and
(\ref{eq:e1kxe0ye0})-(\ref{eq:e3kxe0ye0}) for different cases
of $k_x$ and $k_y$, are used to find $e_1(\vec{k})$,
$e_2(\vec{k})$, and $e_3(\vec{k})$ from given $s_x(\vec{i})$ and
$s_y(\vec{i})$, which lead to $e_1(\vec{i})$, $e_2(\vec{i})$, and
$e_3(\vec{i})$ configurations.
Equations (\ref{eq:uxnFe123})-(\ref{eq:uyxe0ye0}) are used to find the displacements,
$u_x(\vec{i})$ and $u_y(\vec{i})$, from the distortion modes.

\section{Gradients of the energy expression for simulations using Euler method}
The gradient of $E'_{tot}(s_x,s_y)$ necessary for the Euler method is
found from
\begin{eqnarray}
\frac{\partial E'_{tot}(s_x,s_y)}{\partial s_x(\vec{i})}  &=&
 \frac{\partial (E_{l})_{k_x\neq0,k_y\neq0}} {\partial s_x(\vec{i})}+
 \frac{\partial (E_{c})_{k_x\neq0,k_y\neq0}} {\partial s_x(\vec{i})}
 + \frac{\partial (E_l+E_c)^{min}_{k_x=0,k_y\neq0}}{\partial s_x(\vec{i})}
 \nonumber \\
& &+ \frac{\partial (E_l+E_c)^{min}_{k_x\neq0,k_y=0}}{\partial s_x(\vec{i})}
 + \frac{\partial (E_l+E_c)^{min}_{k_x=0,k_y=0} }{\partial s_x(\vec{i})}
 + \frac{\partial E_s}{\partial s_x(\vec{i})},
\end{eqnarray}
\begin{eqnarray}
\frac{\partial E'_{tot}(s_x,s_y)}{\partial s_y(\vec{i})}  &=&
 \frac{\partial (E_{l})_{k_x\neq0,k_y\neq0}} {\partial s_y(\vec{i})} +
 \frac{\partial (E_{c})_{k_x\neq0,k_y\neq0}} {\partial s_y(\vec{i})}
 + \frac{\partial (E_l+E_c)^{min}_{k_x=0,k_y\neq0}}{\partial s_y(\vec{i})}
 \nonumber  \\
& &
 + \frac{\partial (E_l+E_c)^{min}_{k_x\neq0,k_y=0}}{\partial s_y(\vec{i})}
+ \frac{\partial (E_l+E_c)^{min}_{k_x=0,k_y=0}}{\partial s_y(\vec{i})}
 + \frac{\partial E_s}{\partial s_y(\vec{i})}.
\end{eqnarray}
The expression for each term is given below.
\begin{eqnarray}
 \frac{\partial (E_{l})_{k_x\neq0,k_y\neq0}} {\partial s_x(\vec{i})}&=&
\sum_{k_x\neq0,k_y\neq0} [{B}_{xx}(\vec{k})s_x(\vec{k})+{B}_{xy}(\vec{k})s_y(\vec{k})] e^{i\vec{k}\cdot\vec{i}}
\\
 \frac{\partial (E_{l})_{k_x\neq0,k_y\neq0}} {\partial s_y(\vec{i})}&=&
\sum_{k_x\neq0,k_y\neq0} [{B}_{yy}(\vec{k})s_y(\vec{k})+{B}_{xy}(\vec{k})s_x(\vec{k})] e^{i\vec{k}\cdot\vec{i}}
\\
\frac{\partial (E_{c})_{k_x\neq0,k_y\neq0}} {\partial s_x(\vec{i})}&=&
-C_3 \sum_{k_x\neq0,k_y\neq0} \frac{-i}{\sqrt{2}} \cot \frac{k_y}{2} \mathcal{F}_{\vec{k}}(s_x^2-s_y^2) e^{i\vec{k}\cdot\vec{i}}
\nonumber \\
& &+ 2 C_3 s_x(\vec{i})  \sum_{k_x\neq0,k_y\neq0}
\frac{-i}{\sqrt{2}}\left[\cot \frac{k_y}{2} s_x(\vec{k})-\cot \frac{k_x}{2} s_y(\vec{k})\right] e^{i\vec{k}\cdot\vec{i}}
\end{eqnarray}
\begin{eqnarray}
\frac{\partial (E_{c})_{k_x\neq0,k_y\neq0}} {\partial s_y(\vec{i})}&=&
-C_3 \sum_{k_x\neq0,k_y\neq0} \frac{-i}{\sqrt{2}} \cot \frac{k_x}{2} \mathcal{F}_{\vec{k}}(s_y^2-s_x^2) e^{i\vec{k}\cdot\vec{i}}
\nonumber \\
& &+ 2 C_3 s_y(\vec{i})  \sum_{k_x\neq0,k_y\neq0}
\frac{-i}{\sqrt{2}}\left[\cot \frac{k_x}{2} s_y(\vec{k})-\cot \frac{k_y}{2} s_x(\vec{k})\right] e^{i\vec{k}\cdot\vec{i}}
\\
\frac{\partial (E_l+E_c)^{min}_{k_x=0,k_y\neq0}}{\partial s_x(\vec{i})}&=&
  -\frac{2C_3^2}{A_1+A_3}s_x(\vec{i}) \sum_{k_x=0,k_y\neq0} e^{i k_y i_y} \mathcal{F}_{0,k_y}(s_x^2-s_y^2)
\\
 \frac{\partial (E_l+E_c)^{min}_{k_x=0,k_y\neq0}}{\partial s_y(\vec{i})}&=&
  \frac{2C_3^2}{A_1+A_3} s_y(\vec{i}) \sum_{k_x=0,k_y\neq0} e^{i k_y i_y} \mathcal{F}_{0,k_y}(s_x^2-s_y^2)
\\
\frac{\partial (E_l+E_c)^{min}_{k_x\neq0,k_y=0}}{\partial s_x(\vec{i})} &=&
 -\frac{2C_3^2}{A_1+A_3} s_x(\vec{i}) \sum_{k_x\neq0,k_y=0} e^{i k_x i_x} \mathcal{F}_{k_x,0}(s_x^2-s_y^2)
\\
\frac{\partial (E_l+E_c)^{min}_{k_x\neq0,k_y=0}}{\partial s_y(\vec{i})} &=&
 \frac{2C_3^2}{A_1+A_3} s_y(\vec{i}) \sum_{k_x\neq0,k_y=0} e^{i k_x i_x} \mathcal{F}_{k_x,0}(s_x^2-s_y^2)
\\
\frac{\partial (E_l+E_c)^{min}_{\vec{k}=0}}{\partial s_x(\vec{i})} &=&
 -\frac{2C_3^2}{A_3} s_x(\vec{i}) \mathcal{F}_{\vec{k}=0}(s_x^2-s_y^2)
\\
\frac{\partial (E_l+E_c)^{min}_{\vec{k}=0}}{\partial s_y(\vec{i})} &=&
 \frac{2C_3^2}{A_3} s_y(\vec{i})  \mathcal{F}_{\vec{k}=0}(s_x^2-s_y^2)
\\
 \frac{\partial E_s}{\partial s_x(\vec{i})} &=& B s_x(\vec{i})+ G_1 s_x(\vec{i})^3 + G_2 s_x(\vec{i}) s_y(\vec{i})^2
+ H_1 s_x(\vec{i})^5 \nonumber
\\
& &+\frac{H_2}{3} \left[ 2 s_x(\vec{i})^2 + s_y(\vec{i})^2 \right] s_x(\vec{i}) s_y(\vec{i})^2 \\
 \frac{\partial E_s}{\partial s_y(\vec{i})}&=&B s_y(\vec{i})+ G_1 s_y(\vec{i})^3 + G_2 s_x(\vec{i})^2 s_y(\vec{i})
+ H_1 s_y(\vec{i})^5  \nonumber \\
& &+\frac{H_2}{3} \left[ 2 s_y(\vec{i})^2 + s_x(\vec{i})^2 \right] s_y(\vec{i}) s_x(\vec{i})^2
\end{eqnarray}

\newpage

\begin{thebibliography}{99}
\bibitem{Anisotropy}
We note that because of the anisotropy, e.g. $cos(4\theta)$ in 2D in Ref.~\onlinecite{Shenoy99},
the interaction is more convergent than for $1/r^D$.
\bibitem{Final}
Small mistake has been found in the code used in Ref.~\onlinecite{Ahn04},
which was corrected here, responsible for minor differences between
Figs.~\ref{fig:shallow1}(a)-\ref{fig:shallow1}(g)
and the corresponding figures in Ref.~\onlinecite{Ahn04}.
The difference in Fig.~\ref{fig:shallow1}(h)
is because the simulation in Ref.~\onlinecite{Ahn04} was stopped
when the energy change is considered as sufficiently small, not when
the energy change fluctuates, which is used as condition for the final stability
in current Euler simulations.
\end{thebibliography}
\end{document}